\documentclass[aps,superscriptaddress,amsmath,amssymb,floatfix,twocolumn,showpacs,amsfonts,longbibliography]{revtex4-1}
\usepackage{times}
\usepackage[varg]{txfonts}
\usepackage{textcomp}
\usepackage{graphicx}
\usepackage{subfigure}
\usepackage{color}
\usepackage[colorlinks=true,citecolor=blue,urlcolor=blue,linkcolor=blue,hyperindex]{hyperref}
\usepackage{braket}
\usepackage{overpic}
\usepackage{bm}
\usepackage{ulem}

\def\pp{\mathbf{p}}
\def\kk{\mathbf{k}}
\def\qq{\mathbf{q}}
\def\bs{\mathbf{S}}

\def\bq{\mathbf{Q}}
\def\bdelta{\boldsymbol{\delta}}
\def\non{\nonumber}

\begin{document}

\title{Torque equilibrium spin wave theory of Raman scattering in an anisotropic triangular lattice antiferromagnet with Dzyaloshinskii-Moriya interaction}
\author{Chao Shan}
\thanks{These two authors contributed equally.}
\affiliation{State Key Laboratory of Optoelectronic Materials and Technologies, School of Physics, Sun Yat-Sen University, Guangzhou 510275, China}
\author{Shangjian Jin}
\thanks{These two authors contributed equally.}
\affiliation{State Key Laboratory of Optoelectronic Materials and Technologies, School of Physics, Sun Yat-Sen University, Guangzhou 510275, China}

\author{Trinanjan Datta}
\email[Corresponding author:]{tdatta@augusta.edu}
\affiliation{Department of Chemistry and Physics, Augusta University, 1120 15th Street, Augusta, Georgia 30912, USA}
\affiliation{State Key Laboratory of Optoelectronic Materials and Technologies, School of Physics, Sun Yat-Sen University, Guangzhou 510275, China}

\author{Dao-Xin Yao}
\email[Corresponding author:]{yaodaox@mail.sysu.edu.cn}
\affiliation{State Key Laboratory of Optoelectronic Materials and Technologies, School of Physics, Sun Yat-Sen University, Guangzhou 510275, China}

\date{\today}
\begin{abstract}
We apply torque equilibrium spin wave theory (TESWT) to investigate an anisotropic XXZ antiferromagnetic model with Dzyaloshinskii-Moriya (DM) interaction in a triangular lattice. Considering the quasiparticle vacuum as our reference, we provide an accurate analysis of the non-collinear ground state of a frustrated triangular lattice magnet using the TESWT formalism. We elucidate the effects of quantum fluctuations on the ordering wave vector based on model system parameters. We study the single magnon dispersion, the two-magnon continuum using the spectral function, and the Raman spectrum of bimagnon and trimagnon excitations. We present our results for the $HH, VV$, and the $HV$ polarization Raman geometry dependence of the bimagnon and the trimagnon excitation spectrum where $H (V)$ represents horizontal (vertical) polarization. Our calculations show that both the $HH$ and the $HV$ polarization spectrum can be used to determine the degree of anisotropy of our system. We calculate the Raman spectra of Ba$_3$CoSb$_2$O$_9$ and Cs$_2$CuCl$_4$.
\end{abstract}
\maketitle
\section{Introduction}\label{sec:intro}
The effect of quantum fluctuations on the ground state and the phase diagram of frustrated magnets has been a topic of interest in recent years~\cite{nphys749,PhysRevB.79.184413,Hauke_2011,PhysRevB.84.245130,PhysRevB.87.165123,PhysRevB.90.184414,SchmidtPhysRevB.89.184402,StarykhPhysRevLett.113.087204}. The two-dimensional triangular lattice antiferromagnet (TLAF) is a canonical example of a frustrated magnetic system. There are several examples of TLAF, see Table~\ref{Tab1}. Due to the melting of long-range magnetic order by
frustration~\cite{nature08917,b96825}, the two-dimensional triangular lattice is considered as a natural spin liquid candidate~\cite{doi:10.1146/annurev-conmatphys-062910-140521,Powell_2011}. Even when magnetic long-range-order exists, the competition between various interactions have consequences on the ground state and the phase diagram, especially for low-dimensional spin systems~\cite{PhysRevB.92.214409,PhysRevB.94.134416}. Quantum fluctuations can be non-negligible even for ordered magnets with 120$^{\circ}$ spiral order~\cite{PhysRevLett.116.087201,PhysRevLett.109.127203,PhysRevB.96.024416}.

The non-collinear spin structure of the triangular lattice leads to interesting phenomena such as the presence of a roton minimum~\cite{PhysRevLett.96.057201,PhysRevB.74.224420} and a continuum of high-energy magnons~\cite{PhysRevB.68.134424,PhysRevLett.109.267206,PhysRevLett.111.257202}. Similar to superfluid $^4$He~\cite{book:780611} and fractional quantum Hall systems~\cite{PhysRevB.33.2481}, Zheng~\emph{et. al.}~\cite{PhysRevB.68.134424,PhysRevLett.109.267206,PhysRevLett.111.257202} defined the $M$ and $M^\prime$ points of the Brillouin zone (BZ) of a triangular lattice as rotonlike points. The formation of the local minimum is caused by quantum fluctuations \cite{PhysRevB.90.014421,PhysRevLett.115.207202}. The roton signal has been observed in inelastic neutron scattering (INS) experiments~\cite{PhysRevLett.109.127203,PhysRevLett.96.057201,PhysRevB.74.180403,PhysRevB.79.144416}. However, the nature of the high-energy continuum in the triangular lattice is still controversial. The continuous excitation at high energy~\cite{PhysRevLett.109.267206,PhysRevLett.111.257202,PhysRevB.68.134424} may come from fractional excitation of a proximate spin-liquid phase~\cite{PhysRevB.72.174417,PhysRevLett.95.247203,PhysRevB.74.014408,PhysRevLett.98.077205,PhysRevB.91.134423} or from strong magnon-magnon interactions~\cite{PhysRevB.72.134429,PhysRevB.73.184403,PhysRevB.88.094407}. In this context, Raman spectroscopy serves as a powerful tool to probe lattice distortions and the effect of ground state quantum fluctuations. It has already been used to detect magnon excitations in  TLAF~\cite{Sugawara_1993,Suzuki_1993,Aktas_2011,Wulferding_2012,PhysRevB.91.144411,PhysRevB.92.161112,2005.06073}. Raman's advantage is the sensitivity to polarization geometry~\cite{Vernay_2007} and magnon-magnon interactions~\cite{PhysRevB.77.174412,PhysRevB.87.174423}, which is helpful for studying the high energy continuum. Note, a previous resonant inelastic x-ray scattering (RIXS) calculation~\cite{PhysRevB.100.054410} on an anisotropic TLAF has investigated the polarization-independent bi- and tri- magnon spectrum.

There are a couple of Raman scattering experiments on the anisotropic triangular lattice compounds, $\alpha-$CaCr$_{2}$O$_{4}$~\cite{Wulferding_2012} and  $\alpha-$SrCr$_{2}$O$_{4}$ ~\cite{PhysRevB.91.144411}. However, the model Hamiltonian for these two compounds is complicated. From a theoretical perspective, Raman bimagnon calculation for the TLAF has been performed with a square lattice Raman scattering operator~\cite{PhysRev.166.514} within the framework of interacting spin wave theory~\cite{PhysRevB.77.174412, PhysRevB.87.174423}. However, the presence of divergence in the ordering wave vector and singularity of the spin wave spectrum calls for renewed attention to accurately describe the non-collinear frustrated triangular lattice magnet~\cite{PhysRevB.100.054410} beyond the 1/$S$-spin wave theory analysis. The Raman spectrum should be carefully reconsidered with appropriate quantum fluctuation effects and with the proper underlying lattice symmetry. The recently established torque equilibrium spin wave theory (TESWT) considers the spin Casimir effect of a non-collinear system caused by the zero-point quantum fluctuations~\cite{PhysRevB.92.214409}. This formalism cures the ordering wave vector of any divergence and is as convenient as 1/S-spin wave theory. The computed phase diagram of the anisotropic TLAF is consistent with series expansion (SE) and modified spin wave theory (MSWT)~\cite{Hauke_2011,PhysRevB.59.14367}. Since quantum fluctuations cause modification of the ordering wave vector, its influence on the magnon and multi-magnon excitations (bi- and tri-magnon) is an important question to investigate. 

\begin{table}[t]
    \centering
	\caption{Ordered antiferromagnetic triangular lattice materials. The third (fourth) column is the nearest-neighbor exchange interaction $J$ (ordering wave vector). Our torque equilibrium spin wave theory approach for Raman spectrum calculation can be applied to any of the ordered triangular lattice materials listed below. In this article, we only report on the Raman  scattering spectrum of Ba$_3$CoSb$_2$O$_9$ and Cs$_2$CuCl$_4$.}
\begin{tabular}{p{70pt}p{60pt}p{40pt}p{60pt}}
\hline
\hline
Material &Space group &$J$ (meV) &$\bq$ \\
\hline
Ba$_3$CoSb$_2$O$_9$~\cite{PhysRevLett.110.267201} &P6$_{3}$/mmc &1.67 &(2/3,0,1)\\
CuCrO$_2$~\cite{PhysRevB.84.094448} &R$\bar{3}$m &2.8 &(0.658,0,0)\\
$\alpha$-SrCr$_2$O$_4$~\cite{PhysRevB.96.024416} &Pmmn &$J_{mean}\approx5$ &(0.6609,0,1)\\
$\alpha$-GaCr$_2$O$_4$~\cite{PhysRevLett.109.127203} &Pmmn &8.8 &(0.6659,0,1)\\
LuMnO$_3$~\cite{PhysRevLett.111.257202} &P6$_3$cm &9 &(2/3,0,0)\\
Cs$_2$CuCl$_4$~\cite{PhysRevB.100.054410} &Pnma &0.48 &(0.530,0,0)~\cite{PhysRevB.68.134424}\\
Cs$_2$CuBr$_4$~\cite{PhysRevLett.112.077206} &Pnma &1.35 &(0.575,0,0)\\
\hline
\hline
\end{tabular}
\label{Tab1}
\end{table}

In this study, we extend the analysis of the $J$-$J^\prime$ triangular lattice Heisenberg magnet to the case of a XXZ model with DM interaction. First, we apply TESWT to obtain the ordering wave vector. We find that the DM interaction is more favorable to stabilizing the helix state compared to XXZ anisotropy. Similar to the quasi-one-dimensional helimagnets~\cite{PhysRevB.94.134416}, the mere presence of XXZ anisotropy can lead to a shift in the phase boundary. Second, we calculate the spectral function within the TESWT framework. We find that the magnon excitations are more stable with DM interaction and XXZ anisotropy. Our calculations, show the presence of quasiparticle excitation and continuum in the spectral function. Third, we calculate the polarization-dependent bi- and trimagnon Raman spectrum under TESWT. Distinct from the non-interacting calculation, the bimagnon spectrum in the $HV$ polarization displays a single peak feature with magnon-magnon interactions considered. We find that the bimagnon intensity is polarization-independent for the isotropic TLAF. However, the bimagnon excitation occurs only in $HV$ polarization for the anisotropic TLAF. In the $HV$ polarization, spatial anisotropy reduces the bimagnon intensity and peak energy. DM interaction also reduces its intensity, especially in a system with increasing spatial anisotropy. In spite of a spin gap, DM interaction induces an upshift of the bimagnon peak towards higher energy, while XXZ anisotropy downshifts the bimagnon peak to a slightly lower energy. The trimagon excitation is considerable and contributes to the continuum in the $HH$ polarization for the TLAF. We also compute the Raman spectrum of Ba$_3$CoSb$_2$O$_9$ and Cs$_2$CuCl$_4$. For Ba$_3$CoSb$_2$O$_9$, its Raman spectrum has a sharp peak and a broad shoulder in both the $HH$ and the $HV$ polarization. For Cs$_2$CuCl$_4$, its Raman spectrum has a sharp bimagnon peak in $HV$ polarization. We find that the primary contribution to the Raman intensity in the $HH$ polarization comes from the trimagnon excitation. However, in the $HV$ polarization, both the bimagnon and the trimagnon excitation mix and give a broad spectrum.

This article is organized as follows. In Sec.~\ref{sec:model} we introduce the XXZ model with spatial anisotropy and DM interaction. In Sec.~\ref{sec:spin wave} we compute the spin wave spectrum and spectral function by applying TESWT. In Sec.~\ref{sec:raman} we utilize TESWT to calculate the bimagnon and trimagnon Raman spectrum. In Sec.~\ref{subsec:ramantheo} we derive the expressions for the Raman operator, their polarization dependence, and the magnon-magnon interaction effects. In Sec.~\ref{subsec:tuning} we present and discuss our results on spatial anisotropy, spin anisotropy, magnon-magnon interaction, and polarization dependence. Then we present the Raman spectrum of Ba$_3$CoSb$_2$O$_9$ and Cs$_2$CuCl$_4$. Finally, in Sec.~\ref{sec:conclu} we provide our conclusions

\section{MODEL}\label{sec:model}
\begin{figure}[t]
\centering
{
\subfigure[]{
\includegraphics[width=1.55in]{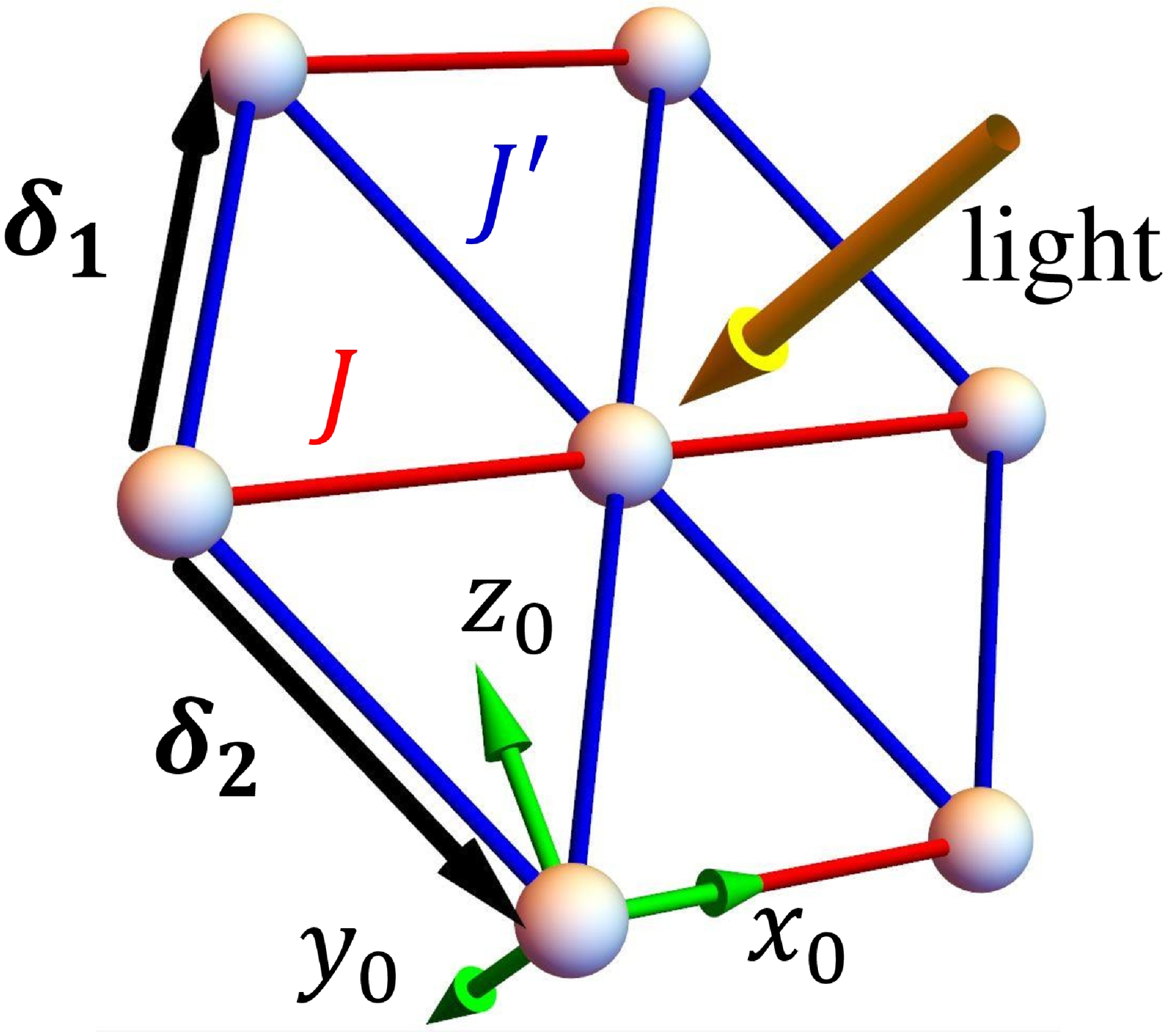}\label{fig:fig1a}}
}
{
\subfigure[]{
\includegraphics[width=1.65in]{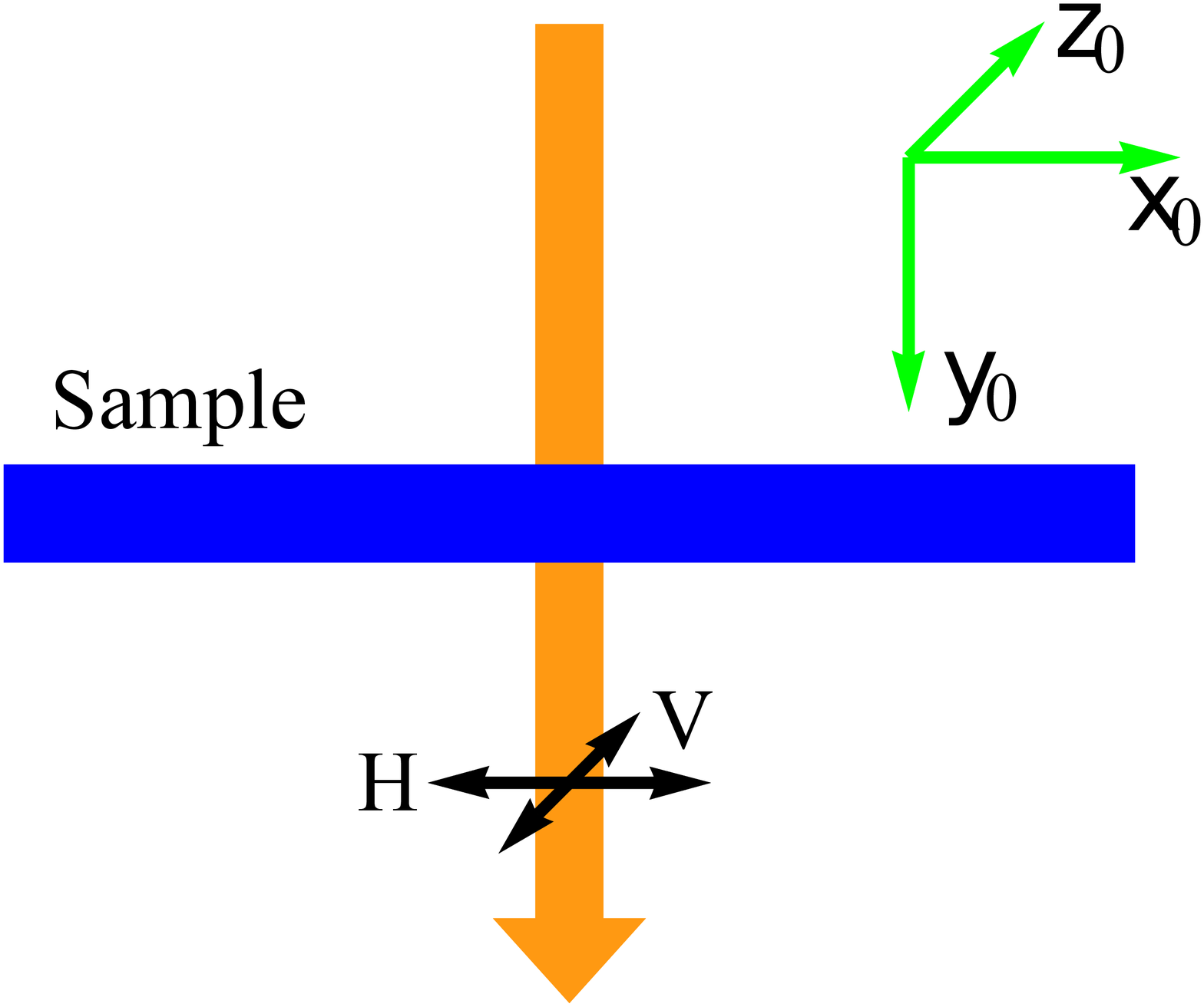}\label{fig:fig1b}}
}
\caption{(a) Triangular lattice with anisotropic exchange constants $J$ and $J^{'}$ acting along the bonds. (b) Experimental geometry setup to study polarization effects within Raman scattering. $\bdelta_{1,2}$ denotes lattice vectors. $H$ (horizontal) and $V$ (vertical) indicates polarization direction of the incoming and outgoing light.}
\label{Fig1}
\end{figure}

Triangular lattice materials can contain spatial or spin anisotropies. In the case of Cs$_2$CuCl$_4$ and Cs$_2$CuBr$_4$, DM interaction is present and generates a spin gap. However, the large gap of Cs$_2$CuBr$_4$ in the energy dispersion cannot be generated exclusively by the DM interaction~\cite{PhysRevLett.112.077206,Zvyagin_2015}. Additionally, spin-orbit coupling may lead to XXZ anisotropy which has been used to explain the presence of a gapped spectrum in some TLAF materials~\cite{PhysRevB.68.104426,PhysRevLett.108.057205}. Even though Ba$_3$CoSb$_2$O$_9$ is spatially isotropic in its exchange interaction, its magnetization is well explained by a spin-1/2 XXZ (spin-anisotropic) model on a triangular lattice~\cite{PhysRevLett.110.267201,PhysRevB.91.024410,PhysRevLett.112.127203,PhysRevLett.112.259901,PhysRevLett.114.027201}. Thus, to conduct a thorough study of the frustrated TLAF systems, we consider XXZ anisotropy in addition to anisotropic exchange interaction and DM interaction. Our model is written as \begin{align}\label{Eq:hammodel}
  \mathcal{H}=&J\sum_{\langle ij\rangle}^{\bdelta_1+\bdelta_2}\Big[S_i^x S_j^x+S_i^z S_j^z+\Delta S_i^y S_j^y\Big]\non\\
  &+J^\prime\sum_{\langle ij\rangle}^{\bdelta_1,\bdelta_2}\Big[S_i^x S_j^x+S_i^z S_j^z+\Delta S_i^y S_j^y\Big]\non\\
  &-\sum_{\langle ij\rangle}^{\bdelta_1,\bdelta_2}\mathbf{D}\cdot(\bs_i\times\bs_j),
\end{align}
where $\langle ij\rangle$ refers to nearest-neighbor bonds on the triangular lattice and $\bdelta_{1,2}$ are the nearest-neighbor (nn) vectors along the diagonal bonds, see Fig.~\ref{fig:fig1a}. The four parameters $(J,J^\prime,D,\Delta)$ contained in the model correspond to the exchange constants along the horizontal bonds, the exchange constants along the diagonal bonds, DM interaction along the $y_0$ direction ($D\textgreater0$)~\cite{PhysRevB.68.134424}, and XXZ spin anisotropy, respectively. 

The spin spiral ground state can be described by an ordering wave vector $\bq$. To analyze the spin wave spectrum of this magnetic model, we first transform from the lab to the rotated local coordinate frame~\cite{PhysRevB.100.054410}. Then, successive applications of the Holstein-Primakoff (HP), the Fourier, and the Bogoliubov transformations give us the effective first-order 1/$S$ expansion Hamiltonian as \begin{eqnarray}\label{Eq:effectiveham}
 \mathcal{H}_{\mathrm{eff}}=&&\sum_\kk\Big[(S\varepsilon_\kk+\delta\varepsilon_\kk)c_\kk^\dag c_\kk+\frac{O_\kk}{2}(c_\kk^\dag c_{-\kk}^\dag+c_\kk c_{-\kk})\Big]\non\\
 &&+\Big\{i\sqrt{\frac{S}{2N}}\Big[\frac{1}{2!}\sum_{\{\kk_i\}}\Phi_{a}(1,2;3)c_{1}^\dag c_{2}^\dag c_{3}\non\\
 &&+\frac{1}{3!}\sum_{\{\kk_i\}}\Phi_{b}(1,2,3)c_{1}^\dag c_{2}^\dag c_{3}^\dag\Big]+\mathrm{H.c.}\Big\}\non\\
 &&+\frac{1}{8N}\sum_{\{\kk_i\}}\Phi_{c}(1,2;3,4)c_{1}^\dag c_{2}^\dag c_{3}c_{4},
\end{eqnarray}
where $c_\kk^\dag$ ($c_\kk$) is the quasiparticle creation (annihilation) operator in momentum space. The numbers 1, 2, 3, ... denote the wave vectors $\kk_1, \kk_2, \kk_3, ...$. The formulae of the vertex coefficients $\Phi_{a}$, $\Phi_{b}$ and $\Phi_{c}$ are given in the Appendix. We set $J=1$ meV in all our subsequent calculations. Thus, our model has three parameters $(J^\prime,D,\Delta)$, where $J^\prime$ and $D$ represent interaction strength relative to the exchange interaction $J$. In the next section, we will analyze the above bosonized Hamiltonian for its spin wave spectrum. As mentioned before, the 1/$S$-spin wave theory cannot treat the quantum fluctuations appropriately, which in turn leads to divergences and singularities in the calculation of the ground state~\cite{PhysRevB.94.134416}. Thus, we will analyze this model using TESWT.

\section{TESWT Spin Wave Spectrum}\label{sec:spin wave}
\begin{figure}[t]
\centering\includegraphics[width=0.48\textwidth]{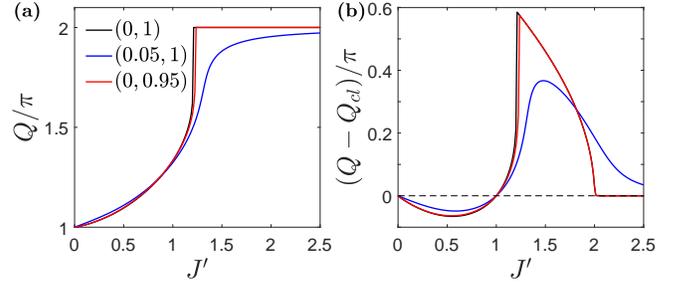}
\caption{Ordering wave vector (a) $Q/\pi$ and (b) $(Q-Q_{cl})/\pi$ versus $J^\prime$ for spin-1/2 system. The black, blue, and red lines show  results for parameters $(D,\Delta)$ equal to $(0,1)$, $(0.05,1)$ and $(0,0.95)$, respectively.}
\label{Fig2}
\end{figure}

Torque equilibrium spin wave theory formalism gives the correct ground state and phase diagram for spin spiral magnets~\cite{PhysRevB.92.214409,PhysRevB.94.134416,PhysRevB.100.054410}. The phase diagram that results from TESWT formalism is consistent with previous numerical calculations~\cite{Hauke_2011,PhysRevB.59.14367}. The essential conceptual difference between spin wave theory and TESWT is the reference ground state. The former considers the classical vacuum as the ground state, while the later considers the quasiparticle vacuum state as the correct starting point. Though linear spin wave spectrum $\varepsilon_\kk$ is physically well-behaved at the classical ordering wave vector $\bq_{cl}=(Q_{cl},0,0)$, it yields an incorrect ground state wave vector, see Fig.~\ref{Fig2}. Within the TESWT approach the goal is to find a redefined Hamiltonian whose classical ordering wave vector $\widetilde{\bq}_{cl}$ is equal to the final ordering wave vector $\bq$ of the original state. Henceforth, the tilde variable will signify parameters of the torque equilibrium shifted Hamiltonian.

To implement TESWT we rewrite the quadratic term of our model as $H_2(J^\prime,D,\Delta,\bq)=\widetilde{H}_2(\widetilde{J}^\prime,\widetilde{D},\widetilde{\Delta},\bq)+H_2^c$, where the superscript $c$ represents the counterterm which will regularize the original singular Hamiltonian. Due to the small values of $D$ and $(1-\Delta)$ in real materials, we take $\widetilde{D}=D$ and $\widetilde{\Delta}=\Delta$. Next, the spin Casimir torque is defined as
\begin{equation}
  \mathbf{T}_{sc}(\bq) = \sum_{\kk}{\Bigg\langle\Psi_{vac}\left| \frac{\partial H_{sw}}{\partial \bq}\right| \Psi_{vac}}\Bigg\rangle,
\end{equation}
where $|\Psi_{vac}\rangle$ represents the expectation value of the quasiparticle vacuum state. Next, we utilize the torque equilibrium condition, within the approximation of $\mathbf{T}_{sc}(\bq)=\widetilde{\mathbf{T}}_{sc}(\bq)$, to obtain the final ordering wave vector as
\begin{align}\label{Eq:ov}
  \frac{\partial E_0(\bq)}{\partial \bq}+\frac{S}{2}\sum_\kk{\frac{\partial \widetilde{\varepsilon}_{\kk}}{\partial \bq}}=0,
\end{align}
where $\widetilde{F}=F(\widetilde{J}^\prime,\widetilde{D},\widetilde{\Delta},\bq)$ ($F$ is an arbitrary operator here). The corresponding functions are shown in Appendix.

Figure~\ref{Fig2}(a) shows the ordering vector $Q$ of the spin-1/2 system obtained using TESWT. Without DM interaction and XXZ anisotropy, the TLAF orders in an antiferromagnet phase for $J^\prime\geq 1.2$. DM interaction influences the ordering wave vector more than XXZ anisotropy. It enlarges the region of spiral phase. Fig.~\ref{Fig2}(b) shows the difference of ordering vector between TESWT and linear spin wave theory (LSWT). For $J^\prime\leq 1$, TESWT gives a smaller $Q$ (compared to LSWT) and the ground state becomes closer to the ferromagnet. Whereas, with $J^\prime\geq 1$ TESWT gives a larger $Q$ and the ground state will be nearly antiferromagnetic in arrangement. Thus we conclude that the spin Casimir effect induces collinear arrangement of spins. Since we are analyzing a coplanar non-collinear spin configuration we will restrict our XXZ anisotropy values. It is evident from Fig.~\ref{Fig2}(b) that with DM interaction, the difference between TESWT and LSWT becomes smaller, indicating that it weakens quantum fluctuations. The Hamiltonian shift results in the one-loop torque equilibrium effective Hamiltonian given by\begin{align}
 \widetilde{\mathcal{H}}_{\mathrm{eff}}=&\sum_\kk\bigg[(S\widetilde{\varepsilon}_\kk+\delta\widetilde{\varepsilon}_\kk)c_\kk^\dag c_\kk+\frac{\widetilde{O}_\kk}{2}(c_\kk^\dag c_{-\kk}^\dag+c_\kk c_{-\kk})\non\\
 &+S\varepsilon_\kk^c c_\kk^\dag c_\kk+\frac{SO_\kk^c}{2}(c_\kk^\dag c_{-\kk}^\dag+c_\kk c_{-\kk})\bigg]\non\\
 &+\Big\{i\sqrt{\frac{S}{2N}}\Big[\frac{1}{2!}\sum_{\{\kk_i\}}\widetilde{\Phi}_{a}(1,2;3)c_{1}^\dag c_{2}^\dag c_{3}\non\\
 &+\frac{1}{3!}\sum_{\{\kk_i\}}\widetilde{\Phi}_{b}(1,2,3)c_{1}^\dag c_{2}^\dag c_{3}^\dag\Big]+\mathrm{H.c.}\Big\}\non\\
 &+\frac{1}{8N}\sum_{\{\kk_i\}}\widetilde{\Phi}_{c}(1,2;3,4)c_{1}^\dag c_{2}^\dag c_{3}c_{4},
\end{align}
with
\begin{align}
\varepsilon_\kk^c&=\frac{1}{\widetilde{\varepsilon_\kk}}(\widetilde{A}_\kk A_\kk-\widetilde{B}_\kk B_\kk)-\widetilde{\varepsilon}_\kk,\non\\
O_\kk^c&=\frac{1}{\widetilde{\varepsilon_\kk}}(\widetilde{A}_\kk B_\kk-\widetilde{B}_\kk A_\kk).
\end{align}
In such a non-collinear spin system, we consider the renormalization of magnon dispersion up to 1/$S$ order. Thus, the counterterm contributions from $H_3$ and $H_4$ are neglected~\cite{PhysRevB.92.214409,PhysRevB.94.134416}. Within this scheme the first-order renormalized Green's function in the one-loop approximation is given by
\begin{equation}\label{Eq:green}
  G^{-1}(\kk,\omega)=\omega-S\widetilde{\varepsilon}_\kk-\Big[S\varepsilon_\kk^c+\widetilde{\Sigma}_c(\kk)+\widetilde{\Sigma}_3^a(\kk,\omega)+\widetilde{\Sigma}_3^b(\kk,\omega)\Big],
\end{equation}
where $S\varepsilon_\kk^c$ is the counterterm from $H_2$. $\widetilde{\Sigma}_c(\kk)=\delta\widetilde{\varepsilon}_\kk$ describes the quartic correction following mean-field averages. $\widetilde{\Sigma}_3^{a,b}(\kk,\omega)$ are the self-energy contributions from the cubic interaction and are given by
\begin{align}\label{Eq:sigma3}
 &\widetilde{\Sigma}_3^a(\kk,\omega)=\frac{S}{4N}\sum_{\kk_1}\frac{|\widetilde{\Phi}_{a}(\kk_1,\kk-\kk_1;\kk)|^2}{\omega-S\widetilde{\varepsilon}_{\kk_1}-S\widetilde{\varepsilon}_{\kk-\kk_1}+i0^+},\non\\
 &\widetilde{\Sigma}_3^b(\kk,\omega)=-\frac{S}{4N}\sum_{\kk_1}\frac{|\widetilde{\Phi_{b}}(\kk_1,-\kk-\kk_1,\kk)|^2}{\omega+S\widetilde{\varepsilon}_{\kk_1}+S\widetilde{\varepsilon}_{\kk+\kk_1}-i0^+}.
\end{align}
\begin{figure*}[hbtp]
\centering\includegraphics[width=7in]{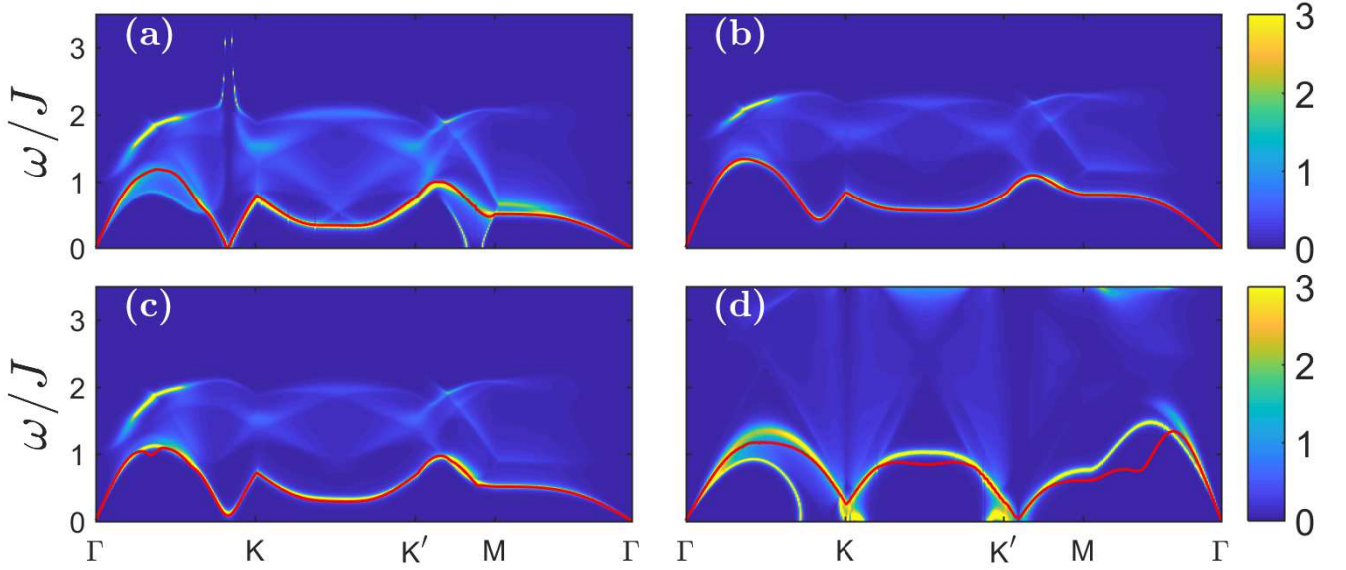}
\caption{Momentum and energy dependence of the spectral function $A({\bf k},\omega)$ for a spin-1/2 system. The points in the chosen path are defined as $\Gamma=(0,0)$, $K=(4\pi/3,0)$, $K^\prime=(2\pi/3,2\pi/\sqrt{3})$ and $M=(0,2\pi/\sqrt{3})$. The parameters $(J^\prime,D,\Delta)$ are (a) (0.5,0,1), (b) (0.5,0.05,1), (c) (0.5,0,0.95) and (d) (1.1,0,1). The solid red lines show the on-shell dispersion $\omega_\kk$, see Eq.~\eqref{Eq:on-shell}.}
\label{Fig3}
\end{figure*}The on-shell approximation for the renormalized Green's function, in the self energy contribution, is  given by $\omega_\kk=S\widetilde{\varepsilon}_\kk$. Thus, the first-order renormalized magnon energy can be calculated as
\begin{equation}\label{Eq:on-shell}
  \omega_\kk=S\widetilde{\varepsilon}_\kk+S\varepsilon_\kk^c+\widetilde{\Sigma}_c(\kk)+\widetilde{\Sigma}_3^a(\kk,S\widetilde{\varepsilon}_\kk)+\widetilde{\Sigma}_3^b(\kk,S\widetilde{\varepsilon}_\kk).
\end{equation} Note, all our derivations are applicable for both low and high spin values. Quantum fluctuations are maximal when S=1/2, as is in our case. The real and imaginary part of $\omega_\kk$ are the magnon dispersion and the magnon decay, respectively. To obtain an intuitive understanding of the single magnon excitation, we calculate the spectral function, which is defined as
\begin{equation}\label{Eq:specfunc}
  A(\kk,\omega)=-\frac{1}{\pi}\mathrm{Im}G(\kk,\omega).
\end{equation}

In Fig.~\ref{Fig3} we report our spectral function calculation for various model parameters. The intensity plots show broadening of the quasiparticle excitation and presence of two-magnon continuum in all the panels, which is consistent with a previous study on the isotropic TLAF~\cite{PhysRevB.88.094407}. Thus, magnon-magnon interactions are important in the triangular lattice. Compared to Ref.~\onlinecite{PhysRevB.88.094407}, spatial anisotropy causes a downshift of the continuum energy. In Fig.~\ref{Fig3}(a) the dispersion shows no gap at the ordering wave vector. This is consistent because the DM interaction is set to zero and the Hamiltonian is at the spin-isotropic point $\Delta=1$.  From Figs.~\ref{Fig3}(b) and~\ref{Fig3}(c) we observe that DM interaction and XXZ anisotropy can suppress damping and stabilize magnon excitations. Both XXZ anisotropy and DM interaction can generate gaps at the ordering  wave vector, thereby reducing magnon decay. However, DM interaction has a greater effect of suppression on magnon decay than XXZ anisotropy. In Fig.~\ref{Fig3}(b), the consistency between the spectral function and dispersion indicates that the on-shell calculation is more reasonable with DM interaction. However, the spectrum is inconsistent with the on-shell dispersion when $J^\prime\textgreater1$, see Fig.~\ref{Fig3}(d). Thus, we restrict our Raman calculations to parameters where the spatially anisotropic exchange interaction does not exceed one. In fact, this is a valid parameter regime for real materials~\cite{PhysRevB.68.104426,PhysRevLett.110.267201,PhysRevLett.112.077206,PhysRevB.100.054410}. 

\section{Torque Equilibrium Spin Wave Theory Raman spectrum}\label{sec:raman}
Raman scattering has the ability to detect magnon excitations due to its sensitivity to magnon-magnon interactions and polarization~\cite{Sugawara_1993,Suzuki_1993,PhysRevB.77.174412,PhysRevB.87.174423,Aktas_2011,Wulferding_2012,PhysRevB.91.144411,Vernay_2007,PhysRevB.95.104431}. Thus, we consider interactions to study the continuum shown in Fig.~\ref{Fig3}. Experimental data from INS indicates that the highest energy of a single magnon is about 20 meV~\cite{PhysRevB.96.024416}. Considering the presence of magnetic interactions in a real material (which are known to introduce downshifts in the Raman spectrum), it is debatable whether the high-energy excitation above 40 meV in the unpolarized Raman experiment~\cite{PhysRevB.91.144411} of $\alpha$-SrCr$_2$O$_4$ may be attributed to a bimagnon. Based on our calculations and the location of the Raman spectrum peaks, we conclude that it most probably contains the trimagnon excitation. Thus, the study of trimagnon Raman spectrum is valuable. In addition to the unpolarized Raman detection, a polarization-dependent Raman spectrum will provide further perspective on the understanding of bi- and tri- magnon excitation. Hence, we study polarized Raman scattering of TLAF to investigate the bi- and tri- magnon excitation behavior. In comparison to RIXS, Raman scattering is a more mature technique restricted to scattering momentum $\qq\approx 0$. For the bimagnon, the momentum of individual magnons can be nonzero, as long as the momenta of the two magnons approximately sum to zero. The same rule applies to the trimagnon.

Till date, from a theoretical perspective, a substantial number of studies have been pursued within LSWT and an interacting framework to investigate the Raman spectrum of TLAF Heisenberg model~\cite{Sugawara_1993,Suzuki_1993,PhysRevB.77.174412,PhysRevB.87.174423}. However, since LSWT leads to a divergent ordering wave vector and fails to describe the ground state, it is not suitable to calculate the magnon excitation. Thus, we apply the TESWT to study the Raman bi- and trimagnon excitation of TLAF. One of the key developments reported in this paper is on trimagnon calculation and our discussion of the polarization dependence of the bi- and tri- magnon excitation. Neglecting polarization, the bimagnon intensity is zero at the $\Gamma$ point. However, the real spectrum of the anisotropic TLAF is polarization dependent. Next, we discuss the polarization dependence and how it helps to analyze the composition of Raman spectrum.

\subsection{Raman scattering operator and interactions}\label{subsec:ramantheo}
Standard perturbation theory formalism applied to electron-radiation interaction can be used to compute the Raman scattering cross-section~\cite{Heitler1954THE,Loudon1964The,1967490}. Since we are studying magnetic Raman scattering, the operator should be expressed in terms of spin operators which obey the underlying lattice symmetry. Our triangular lattice model Hamiltonian contains spatial anisotropy, DM interaction, and XXZ anisotropy. Thus, the expression for the polarization-dependent second-order Raman scattering operator is \begin{align}
\mathcal{\hat{O}}=&\sum_{i,\pm\bdelta_j}\mathcal{P}_j(\theta,\phi)\non\\
&\times\Big[J_j(S_i^x S_{i+\bdelta_j}^x+S_i^z S_{i+\bdelta_j}^z+\Delta S_i^y S_{i+\bdelta_j}^y)-\mathbf{D}_j\cdot(\bs_i\times\bs_{i+\bdelta_j})\Big],
\end{align}
where $\bdelta_j$ denote the lattice vectors: $\bdelta_1=(\frac{1}{2},0,\frac{\sqrt{3}}{2})$, $\bdelta_2=(\frac{1}{2},0,-\frac{\sqrt{3}}{2})$, and $\bdelta_3=(1,0,0)$~\footnote{Note, in our earlier publication Jin~\emph{et al.}~\cite{PhysRevB.100.054410}, there was a typographical error in the reported RIXS scattering operator expression. The expression missed the DM interaction term which was considered in our study. The correct reported form of the RIXS operator expression should be
$\mathcal{R}_\qq=\sum_{i,\bdelta}e^{i\qq\cdot \mathbf{r}_i}[J_{i\bdelta}\bs_i\cdot\bs_{i+\bdelta}-\mathbf{D}_{\bdelta}\cdot(\bs_i\times\bs_{i+\bdelta})]$.}. The polarization geometry and the symmetry of the experimental setup are captured in the $\mathcal{P}_j(\theta,\phi)$ operator coefficient.
We consider the polarization of the incoming and outgoing light as $\hat{\bm{\varepsilon}}_{in}=(\cos\theta,0,\sin\theta)$ and $\hat{\bm{\varepsilon}}_{out}=(\cos\phi,0,\sin\phi)$, respectively, where $\theta$ and $\phi$ are defined with respect to the $x_0$ axis. The sketch of the experimental geometry is shown in Fig.~\ref{fig:fig1b}. Lattice symmetry was employed to analyze the different types of magnetic excitations. Since we are considering a quasi-2D TLAF, for generality, we calculated the Raman spectrum of both the isotropic (C$_{3v}$) and the anisotropic case (C$_{2v}$), respectively. The Raman-active modes of the C$_{3v}$ and C$_{2v}$ systems are given by the irreducible representations $A_1+E$ and $A_1+A_2$, respectively.

In terms of the Bogoliubov magnons the polarized Raman scattering operator takes the following form\begin{align}
\mathcal{\hat{O}}=&\sum_\kk\widetilde{\mathcal{B}}_\kk(c_\kk c_{-\kk}+c_\kk^\dag c_{-\kk}^\dag)\non\\
&+\sum_{\kk,\pp}\widetilde{\mathcal{F}}(\pp,-\kk-\pp,\kk)(c_\pp c_{-\kk-\pp} c_\kk-c_\pp^\dag c_{-\kk-\pp}^\dag c_{\kk}^\dag),
\end{align} where the scattering matrix element $\widetilde{\mathcal{B}}_\kk$ and $\widetilde{\mathcal{F}}(\kk,-\kk-\pp,\pp)$ are given by
\begin{align}
\widetilde{\mathcal{B}}_\kk=S\sum_{j=1}^3 \mathcal{P}_j(\theta,\phi)\Big[\widetilde{\mu}_\kk\widetilde{v}_\kk\xi_{j\kk}-(\widetilde{\mu}_\kk^2+\widetilde{v}_\kk^2)\lambda_{j\kk}\Big],
\end{align} and
\begin{align}
\widetilde{\mathcal{F}}(\kk,-\kk-\pp,\pp)=&\frac{i\sqrt{2S}}{3}\sum_{j=1}^3\mathcal{P}_j(\theta,\phi)\non\\
&\times\Big[\zeta_{j\pp}(\widetilde{\mu}_\pp+\widetilde{v}_\pp)\times(\widetilde{\mu}_{-\kk-\pp}\widetilde{v}_\kk+\widetilde{v}_{-\kk-\pp}\widetilde{\mu}_{\kk})\non\\
&+\zeta_{j,-\kk-\pp}(\widetilde{\mu}_{-\kk-\pp}+\widetilde{v}_{-\kk-\pp})\times(\widetilde{\mu}_{\kk}\widetilde{v}_\pp+\widetilde{v}_{\kk}\widetilde{\mu}_{\pp})\non\\
&+\zeta_{j\kk}(\widetilde{\mu}_{\kk}+\widetilde{v}_{\kk})\times(\widetilde{\mu}_{\pp}\widetilde{v}_{-\kk-\pp}+\widetilde{v}_{\pp}\widetilde{\mu}_{-\kk-\pp})\Big].
\end{align} We note that the summation of the momenta in the bimagnon (trimagnon) scattering matrix element adds up to zero. In the above equations we have introduced the following functions
\begin{align}\label{Eq:introfunction}
\xi_{j\kk}=&2[\Delta J_j+J_j\cos(\bq\cdot\bdelta_j)-D_j\sin(\bq\cdot\bdelta_j)] \cos(\kk\cdot\bdelta_j)\non\\
&- 4[J_j\cos(\bq\cdot\delta_j)-D_j\sin(\bq\cdot\bdelta_j)],\non\\
\lambda_{j\kk}=&[\Delta J_j-J_j\cos(\bq\cdot\bdelta_j)+D_j\sin(\bq\cdot\bdelta_j)] \cos(\kk\cdot\bdelta_j),\non\\
\zeta_{j\kk}=&- [J_j\sin(\bq\cdot\bdelta_j)+D_j\cos(\bq\cdot\bdelta_j)] \sin(\kk\cdot\bdelta_j).
\end{align}
For the C$_{3v}$ symmetry the $\mathcal{P}_j(\theta,\phi)$ coefficient is given by the following function
\begin{align}\label{Eq:c3v}
\mathcal{P}_j(\theta,\phi)=&\varepsilon_{in}(\theta)\left(
\begin{array}{ccc}
p_1\\
&p_2&\\
&&p_1
\end{array}\right)\varepsilon_{out}^\mathrm{T}(\phi)\alpha_j^{A_1}\non\\
&+\varepsilon_{in}(\theta)\left(
\begin{array}{ccc}
p_3\\
&&p_4\\
&p_4&-p_3
\end{array}\right)\varepsilon_{out}^\mathrm{T}(\phi)\alpha_j^{E_1}\non\\
&+\varepsilon_{in}(\theta)\left(
\begin{array}{ccc}
&-p_4&-p_3\\
-p_4\\
-p_3
\end{array}\right)\varepsilon_{out}^\mathrm{T}(\phi)\alpha_j^{E_2},
\end{align}
with $\alpha_j^{A_1}=1$, $\alpha_3^{E_1}=-2\alpha_1^{E_1}=-2\alpha_2^{E_1}=1/2$, $\alpha_3^{E_2}=0$ and $\alpha_1^{E_2}=-\alpha_2^{E_2}=\sqrt{3}/{4}$~\cite{Sugawara_1993,Suzuki_1993}. We note that each irreducible representation contains an overall multiplicative form factor exclusive to that channel. Within the polarization defined above, the scattering spectrum is dependent only on the $(p_1,p_3)$ coefficients. To proceed with the calculation (in the absence of adequate information to calculate these form factors) we make a simplification by setting $p_1=p_3=1$ in all subsequent calculations. Note, $\varepsilon_{out}^\mathrm{T}$ is the transpose of $\varepsilon_{out}$. For the C$_{2v}$ symmetry the $\mathcal{P}_j(\theta,\phi)$ coefficient is given by
\begin{align}\label{Eq:c2v}
\mathcal{P}_j(\theta,\phi)=&\varepsilon_{in}(\theta)\left(
\begin{array}{ccc}
p_5\\
&p_6&\\
&&p_7
\end{array}\right)\varepsilon_{out}^\mathrm{T}(\phi)\eta_j^{A_1}\non\\
&+\varepsilon_{in}(\theta)\left(
\begin{array}{ccc}
&&p_8\\
&0\\
p_8
\end{array}\right)\varepsilon_{out}^\mathrm{T}(\phi)\eta_j^{A_2},
\end{align}
with $\eta_j^{A_1}=1$, $\eta_3^{A_2}=0$, and $\eta_1^{A_2}=-\eta_2^{A_2}=\sqrt{3}/{4}$. Within the defined polarization, the spectrum is independent of the $p_6$ coefficient. To simplify, we set $p_5=p_7=p_8=1$ in our computations.

\begin{figure}[t]
\centering\includegraphics[width=0.48\textwidth]{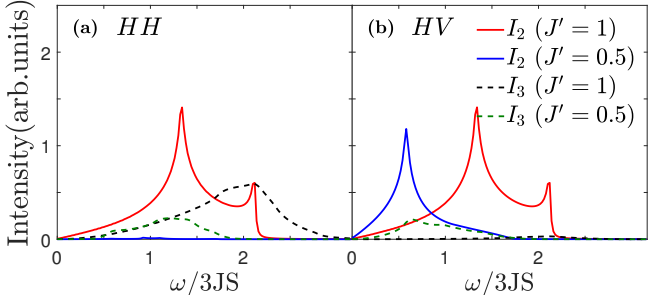}
\caption{Non-interacting bimagnon (solid lines) and trimagnon (dashed lines) Raman spectrum of spin-1/2 system in (a) $HH$ and (b) $HV$ polarization. DM interaction and XXZ anisotropy are absent in the system. The red and black lines are calculated with $J^\prime=1$ under C$_{3v}$ symmetry. The blue and green lines are calculated with $J^\prime=0.5$ under C$_{2v}$ symmetry.}
\label{Fig4}
\end{figure}

According to the fluctuation-dissipation theorem the Raman scattering intensity can be related to the multi-magnon susceptibility. In our particular case, the bi- and tri- magnon susceptibility are defined as
\begin{align}\label{Eq:I2}
\chi_2(\omega)=&\int_0^\beta d \tau e^{iw\tau} \sum_{\kk\kk'}\widetilde{\mathcal{B}}_{\kk}\widetilde{\mathcal{B}}_{\kk^\prime}\langle T_\tau c_{\kk}(\tau)c_{-\kk}(\tau)c_{\kk^\prime}^\dag c_{-\kk^\prime}^\dag\rangle,\\
\chi_3(\omega)=&\int_0^\beta d \tau e^{iw\tau} \sum_{\kk\kk'\pp\pp'}\widetilde{\mathcal{F}}_{\kk,\pp}\widetilde{\mathcal{F}}_{\kk^\prime,\pp^\prime}\non\\
&\langle T_\tau c_{\kk}(\tau)c_{-\kk-\pp}(\tau)c_{\pp}(\tau)c_{\kk^\prime}^\dag c_{-\kk^\prime-\pp^\prime}^\dag c_{\pp^\prime}^\dag\rangle,
\end{align}
where $T_\tau$ is the time-ordering operator. $\langle\cdot\rangle$ is the average of the ground state. Here, we study the case of zero temperature $\beta=1/k_B T$. According to Fermi's golden rule, the non-interacting scattering intensity is related to the bare Green's function $\mathrm{G}_0(\kk,\omega)=1/(\omega-\omega_\kk^{(0)}+i0^+)$ with $\omega_\kk^{(0)}=S\widetilde{\varepsilon}_\kk$ in the quasi-particle representation. Applying Wick's theorem, the non-interacting spectrum can be calculated as
\begin{eqnarray}
I_2(\omega)&=&2\sum_{\kk}\widetilde{\mathcal{B}}_{\kk}^2\delta(\omega-\omega_{\kk}^{(0)}-\omega_{\kk}^{(0)}),\non\\
I_3(\omega)&=&6\sum_{\kk,\pp}\widetilde{\mathcal{F}}_{\kk,-\kk-\pp,\pp}^2\delta(\omega-\omega_{\kk}^{(0)}-\omega_{-\kk-\pp}^{(0)}-\omega_{\pp}^{(0)}).
\end{eqnarray}
The non-interacting result is shown in Fig.~\ref{Fig4} and will be discussed in Sec.~\ref{subsec:tuning} along with the interacting case.

Next, we consider the $1/S$ correction to the bimagnon excitation. The two-particle propagator $\Pi_{\kk\kk'}(\omega)$ from Eq.~\eqref{Eq:I2} is given by
\begin{align}
\Pi_{\kk\kk'}(\omega)=2i\int\frac{\mathrm{d}\omega'}{2\pi}\mathrm{G}_{\kk}(\omega+\omega')
\mathrm{G}_{-\kk}(-\omega')\Gamma_{\kk\kk'}(\omega,\omega').\label{Eq:propa}
\end{align}
The vertex function $\Gamma_{\kk\kk'}(\omega,\omega')$ can be computed from the Bethe-Salpeter equation which is expressed as~\cite{PhysRevB.92.035109}
\begin{eqnarray}
\Gamma_{\kk\kk'}(\omega,\omega')=&&\delta_{\kk\kk'}+\sum_{\kk_1}2i\int\frac{\mathrm{d}\omega_1}{2\pi}
\mathrm{G}_{\kk_1}(\omega+\omega_1)\mathrm{G}_{-\kk_1}(-\omega_1)\non\\
&&\times\mathcal{V}^{\mathrm{IR}}_{\kk\kk_1}(\omega^\prime,\omega_1)\Gamma_{\kk_1\kk'}(\omega,\omega_1),\label{Eq:vertex}
\end{eqnarray}
where $\mathcal{V}^{\mathrm{IR}}_{\kk\kk_1}(\omega^\prime,\omega_1)=\mathcal{V}^{\mathrm{3}}_{\kk\kk_1}(\omega^\prime,\omega_1)+\mathcal{V}^{\mathrm{4}}_{\kk\kk_1}$ is the two-particle irreducible vertex. For the Raman process, the scattering momentum $\qq\approx0$, thus leading to the disappearance of vertices $\mathcal{V}_3^{(c)}$ and $\mathcal{V}_3^{(d)}$. Thus, the cubic vertex is calculated as
\begin{align}
\mathcal{V}^{\mathrm{3}}_{\kk\kk_1}(\omega^\prime,\omega_1)=&\frac{S}{(8N)}[\widetilde{\Phi}_{a}(\kk_1,\kk-\kk_1;\kk)\widetilde{\Phi}^\ast_{a}(-\kk,\kk-\kk_1;-\kk_1)\non\\
&\times \mathrm{G}_0(\kk-\kk_1,\omega'-\omega_1)\non\\
&+\widetilde{\Phi}^\ast_{a}(\kk,\kk_1-\kk;\kk_1)\widetilde{\Phi}_{a}(-\kk_1,\kk_1-\kk;-\kk)\non\\
&\times \mathrm{G}_0(\kk_1-\kk,\omega_1-\omega')].
\end{align}
The four-point vertex $\mathcal{V}^{\mathrm{4}}_{\kk\kk_1}$
originating from the quartic Hamiltonian is given for our Raman case as
\begin{align}
\mathcal{V}^{\mathrm{4}}_{\kk\kk_1}=&\frac{1}{4N}\Big[4(A_0+B_0+\frac{1}{2}A_{\kk+\kk_1}+\frac{1}{2}B_{\kk+\kk_1}+\frac{1}{2}A_{\kk-\kk_1}\non\\
&+\frac{1}{2}B_{\kk-\kk_1}-A_\kk-A_{\kk_1})\mu_\kk\mu_{\kk1}v_\kk v_{\kk1}+(A_{\kk-\kk_1}+B_{\kk-\kk_1}\non\\
&+A_{\kk+\kk_1}+B_{\kk+\kk_1}-A_\kk-A_{\kk_1})(\mu_\kk^2\mu_{\kk_1}^2+v_\kk^2v_{\kk_1}^2)-(2B_\kk\non\\
&+B_{\kk_1})(\mu_\kk^2+v_\kk^2)\mu_{\kk_1}v_{\kk_1}-(2B_{\kk_1}+B_\kk)(\mu_{\kk_1}^2+v_{\kk_1}^2)\mu_\kk v_\kk\Big].
\end{align}
Next, considering the appropriate time domain of $\tau\in[-\infty,\infty]$, the expression for the interacting bimagnon Raman intensity is given by
\begin{align}\label{Eq:interI}
I_2(\omega)=-\frac{1}{\pi}\mathrm{Im}\sum_{\mathrm{m,n}}[\mathrm{\hat{\chi}}_{\mathrm{mn}}(\omega)-\mathrm{\hat{\chi}}_{\mathrm{mn}}(-\omega)].
\end{align}
In the above calculation we will assume that the two on-shell magnons are created and annihilated in the intermediate propagators with $\omega'\approx-\omega_\kk^{(0)}=-S\widetilde{\varepsilon}_\kk-S\varepsilon_\kk^c$ and $\omega_1\approx-\omega_{\kk_1}^{(0)}=-S\widetilde{\varepsilon}_{\kk_1}-S\varepsilon_{\kk_1}^c$. We calculate the bimagnon susceptibility in the matrix form $\mathrm{\hat{\chi}}^T=\widetilde{\mathcal{B}}_\kk\textbf{[}\mathbf{\hat{1}}-\hat{\Gamma}_{\kk,\pp}\textbf{]}^{-1}\Pi_\kk\widetilde{\mathcal{B}}_\kk$, where $\Pi_{\kk}=2[\omega-2\omega_{\kk}+i0^+]^{-1}$ is the renormalized two-magnon propagator~\cite{PhysRevB.92.035109,PhysRevB.100.054410}.

\subsection{Bimagnon and trimagnon Raman spectrum}\label{subsec:tuning}
We calculate the spectrum under $HH, VV$, and $HV$ polarization (as realized in experimental set-up), where $H$ and $V$ represent the horizontal and vertical polarization of the incoming and outgoing light, as shown in Fig.~\ref{Fig1}(b). For example, in our notation $HV$ polarization implies $\hat{\bm{\varepsilon}}_{in}=H$ and $\hat{\bm{\varepsilon}}_{out}=V$. The Raman signal from the isotropic TLAF can be explored by one of these three polarization choices originating from the irreducible representations $\alpha^{A_1}+\alpha^{E_1}$ $(HH,VV)$ and $\alpha^{E_2}$ $(HV)$ modes. Note, $HH = VV$ only for the anisotropic TLAF.

The non-interacting Raman spectra for the isotropic cases are shown in Fig.~\ref{Fig4}. The bimagnon spectrum is polarization-independent since the red solid lines of Fig.~\ref{Fig4}(a) and Fig.~\ref{Fig4}(b) are the same. However, the trimagnon is polarization-dependent since the black dashed lines of Fig.~\ref{Fig4}(a) and Fig.~\ref{Fig4}(b) behave differently. The trimagnon intensity in the $HH$ polarization is much greater than that in the $HV$ polarization. For the anisotropic case, the Raman signal under $HH$ $(VV)$ polarization and $HV$ polarization stems from $\eta^{A_1}$ and $\eta^{A_2}$ mode, respectively. As the Raman scattering operator of the bimagnon in $A_1$ mode is commutable with $H_2$, the bimagnon intensity in this channel is zero, see the blue solid line in Fig.~\ref{Fig4}(a). Different from the two-peak feature observed in the Raman response of the isotropic TLAF in Fig.~\ref{Fig4}, the bimagnon spectrum of the anisotropic case presents a single peak structure with a downshift in peak energy.

\begin{figure}[t]
\centering\includegraphics[width=3.4in]{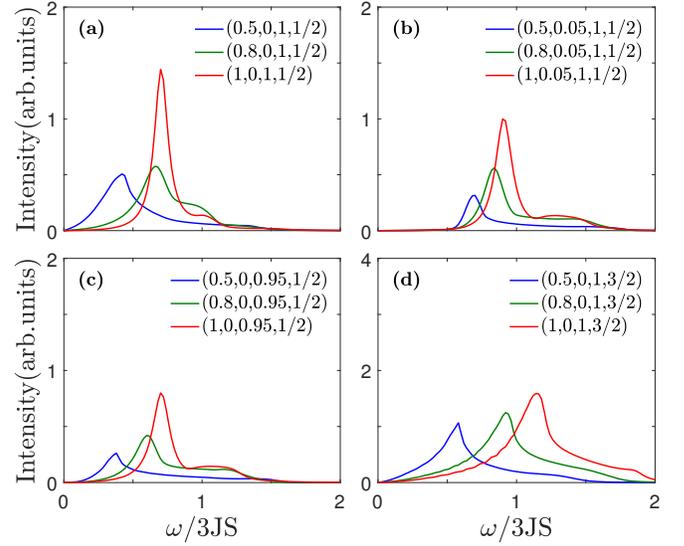}
\caption{Interacting Raman bimagnon spectra of spin-1/2 systems in $HV$ polarization with different parameters $(J^\prime,D,\Delta,S)$.}
\label{Fig5}
\end{figure}
Previously we discussed the importance of magnon-magnon interaction within the context of the spectral function of a TLAF. Thus, we calculated the interacting Raman multi-magnon spectrum using Eq.~\eqref{Eq:interI}. Fig.~\ref{Fig5} shows the interacting bimagnon intensity with different parameters under $HV$ polarization. We note that the bimagnon intensity with $HH$ $(VV)$ polarization is almost zero. We study the effect of magnon-magnon interactions, spatial anisotropy, DM interaction, XXZ anisotropy, and spin value on Raman bimagnon spectrum. Considering interactions, the spectrum in the $HV$ polarization displays a single peak structure in the isotropic model compared to the two-peak structure of the non-interacting calculation. Spatial anisotropy decreases the intensity and the peak energy. DM interaction shifts the peak towards higher energy. Although XXZ anisotropy introduces a gap, the bimagnon peak has a slight downshift in energy.

As anisotropy increases, the system tends to behave like a quasi-1D spin chain. Thus, it reduces the bimagnon intensity and leads to the downshift of the peak similar to what is predicted to occur in the RIXS spectrum ~\cite{PhysRevB.100.054410} at the roton scattering momentum $\qq=M$ and $\qq=M^\prime$. However, unlike the RIXS spectrum, the DM interaction causes a decrease in the Raman intensity, compare Fig.~\ref{Fig5}(a) to Fig.~\ref{Fig5}(b). The reduction effect is also seen in the XXZ model if we compare Fig.~\ref{Fig5}(a) to Fig.~\ref{Fig5}(c). This reduction can be attributed to the influence of DM interaction and XXZ anisotropy on the bimagnon scattering matrix element. With increasing spatial anisotropy, DM interaction plays a more important role. Thus, the Raman intensity reduction is more prominent in the systems with greater spatial anisotropy. The renormalized dispersion is shown in Fig.~\ref{Fig3}. DM interaction introduces a spin gap, resulting in the higher peak energy in Fig.~\ref{Fig5}(b) compared to Fig.~\ref{Fig5}(a). Although a tiny gap is generated by XXZ anisotropy, the bimagnon peak shifts to lower energy slightly in Fig.~\ref{Fig5}(c) compared to Fig.~\ref{Fig5}(a). In addition, the large spin value weakens the quantum fluctuations and magnon-magnon interactions, compare Fig.~\ref{Fig5}(d) to Fig.~\ref{Fig5}(a). Thus, Fig.~\ref{Fig5}(d) shows an energy upshift with vanishing shoulder, similar to the non-interacting case.

\begin{figure}[t]
\centering\includegraphics[width=3.4in]{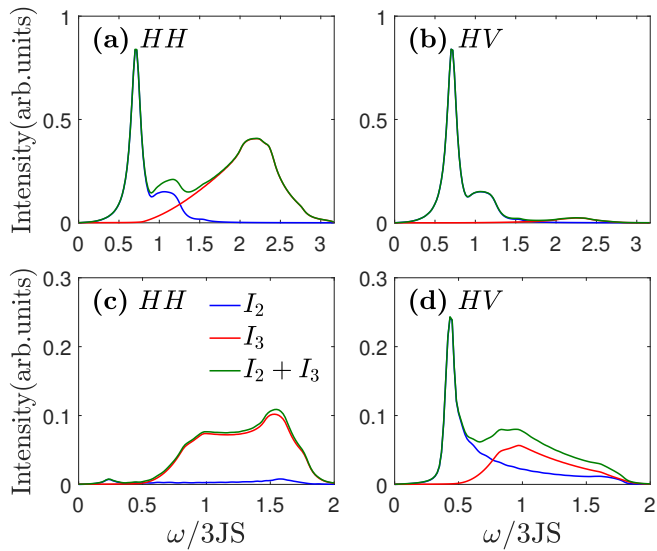}
\caption{Interacting bimagnon and non-interacting trimagnon Raman spectra of Ba$_3$CoSb$_2$O$_9$ and Cs$_2$CuCl$_4$. The left (right) column is under $HH$ ($HV$) polarization. The first line is for Ba$_3$CoSb$_2$O$_9$ with $(J^\prime,D,\Delta)=(1,0,0.954)$~\cite{PhysRevLett.110.267201}. The second line is for Cs$_2$CuCl$_4$ with $(J^\prime,D,\Delta)=(0.316,0.025,1)$~\cite{PhysRevB.100.054410}.}
\label{Fig6}
\end{figure}

We study the Raman spectrum of two real materials Ba$_3$CoSb$_2$O$_9$ (isotropic TLAF) and Cs$_2$CuCl$_4$ (anisotropic TLAF). The Raman spectrum of Ba$_3$CoSb$_2$O$_9$ in $C_{3v}$ symmetry is shown in Fig.~\ref{Fig6}(a) and Fig.~~\ref{Fig6}(b). For Ba$_3$CoSb$_2$O$_9$, with DM interaction set to zero and $J=J^{'}$ the ground state is close to a 120$^{\circ}$ non-collinear magnetic order. We use the parameters from the electronic spin resonance (ESR) experiment of Susuki~\emph{et al.}, Ref.~\onlinecite{PhysRevLett.110.267201}, as an input for our Raman computation. The Raman spectrum of Cs$_2$CuCl$_4$ in $C_{2v}$ symmetry is shown in Fig.~\ref{Fig6}(c) and ~\ref{Fig6}(d). The fit parameters for Cs$_2$CuCl$_4$ were chosen from our earlier TESWT INS fitting result~\cite{PhysRevB.100.054410}.

The Ba$_3$CoSb$_2$O$_9$ Raman spectrum shows two prominent features for both the $HH$ and $HV$ polarization. In the $HH$ polarization there is a clear separation of energy excitation. The bimagnon peaks around $\approx  1.1J$, whereas the trimagnon peaks around $\approx  3.3J$. However, the bimagnon intensity is much greater than the trimagnon response in the $HH$ polarization geometry. For the $HV$ case, the Raman intensity is dominated by the bimagnon signal. For Cs$_2$CuCl$_4$ the bimagnon signal is almost zero in the $HH$ channel. Thus, the trimagnon is the main contribution. This response is different from that observed in the isotropic TLAF Ba$_3$CoSb$_2$O$_9$. Hence, we can use the $HH$ signal as a signature to identify anisotropic behavior in a TLAF. The trimagnon signal is broad and spreads over an energy range of $\approx J- 3J$. In contrast, the $HV$ polarization for Cs$_2$CuCl$_4$ supports a non-zero signal for both the bi- and the trimagnon intensity. The bimagnon peaks at $\approx 0.6J$ and the trimagnon is maximum around $\approx 1.2J$. This behaviour is qualitatively similar to what we observe for Ba$_3$CoSb$_2$O$_9$ Raman signal in the $HH$ geometry. The bimagnon intensity for Ba$_3$CoSb$_2$O$_9$ is greater than Cs$_2$CuCl$_4$. This can be explained by the presence of stronger spin coupling along the diagonal bonds for the isotropic TLAF. Furthermore, comparing Fig.~\ref{Fig6}(b) to Fig.~\ref{Fig6}(d) we find that the trimagnon response survives only in the anisotropic case, consistent with Fig.~\ref{Fig4}(b). Thus, we can also judge the degree of anisotropy of the system from the $HV$ polarized Raman spectrum.
\section{CONCLUSION}\label{sec:conclu}
We applied TESWT to calculate the bi- and the tri- magnon Raman spectrum of an isotropic and an anisotropic TLAF. We extended TESWT to the XXZ model considering both spatial anisotropy and DM interaction. Our calculation is an application of the TESWT formalism to Raman spectroscopy analysis. We computed the TESWT corrected ordering wave vector, the on-shell dispersion, the spectral function, and the Raman spectrum in both $C_{3v}$ and $C_{2v}$ symmetry for the $HH$ and the $HV$ polarization.

Based on our calculations we find that even for our system the spin Casimir effect can induce an arrangement of collinear spins~\cite{PhysRevB.92.214409}. Although DM interaction and XXZ anisotropy stabilize the spiral order, spin wave theory is unable to predict an accurate ordering wave vector for the system. Thus, one needs to account for the presence of spin Casimir torque introduced by zero-point quantum fluctuations. This reduces the range of the spiral phase. Both XXZ anisotropy and DM interaction introduces a gap at the ordering wave vector, leading to suppression of damping and magnon excitation stabilization. We note that DM interaction brings about a stronger suppression effect on magnon decay than XXZ anisotropy. We also discuss the sensitivity of Raman spectrum to polarization, system parameters $(J^\prime,D,\Delta,S)$, and magnon-magnon interactions. We find that large spin values  cause an energy upshift with vanishing shoulder due to the weakened quantum fluctuations and magnon-magnon interactions.
We also compute the bi- and the tri- magnon Raman spectrum for Ba$_3$CoSb$_2$O$_9$ and Cs$_2$CuCl$_4$. In an isotropic TLAF, the Raman spectrum is sensitive to polarization for trimagnon excitation and independent from polarization for the bimagnon excitation. However, the converse holds true for the anisotropic lattice. The bimagnon excitation is polarization dependent. We propose that the degree of anisotropy of the system can be judged using either the $HH$ or the $HV$ polarization. Based on our calculations we have shown that TESWT is a reliable method to calculate and analyze the Raman spectrum of frustrated magnetic materials. Finally, we hope that our results will inspire experimentalists to perform measurements to verify our predictions.

\begin{acknowledgments}
We thank N. Drichko, R. Yu, and Z. Xiong for helpful discussions. T.D. acknowledges invitation, hospitality, and kind support from Sun Yat-Sen University. C.S., S.J., and D.X.Y. are supported by NKRDPC Grants No. 2017YFA0206203, No. 2018YFA0306001, NSFC-11974432, GBABRF-2019A1515011337, and Leading Talent Program of Guangdong Special Projects. T.D. acknowledges funding support from Sun Yat-Sen University Grants No. OEMT--2019--KF--04, No. OEMT--2017--KF--06 and Augusta University Scholarly Activity Award.
\end{acknowledgments}


\appendix
\section*{APPENDIX: $1/S$ spin wave THEORY}
In this Appendix we state the expression for the functions introduced under $1/S$-LSWT, which can be easily applied under the TESWT.
The classical energy $E_0(\bq)$ is given by
\begin{align}
E_0(\bq)=NS^2(J_\bq-\eta_\bq)=NS^2\gamma_\bq,
\end{align}
with
\begin{align}
&J_\kk=J\cos k_x+2J^\prime\cos\frac{k_x}{2}\cos\frac{\sqrt{3}}{2}k_y,\non\\
&\eta_\kk=2D\sin\frac{k_x}{2}\cos\frac{\sqrt{3}}{2}k_y.
\end{align}
The classical ordering vector $Q_{cl}$ is obtained by solving the following self-consistent equation
\begin{equation}\label{Eq:ovL}
\nabla_\bq E_0(\bq)=0.
\end{equation}
The bare magnon dispersion is $S\varepsilon_\kk$ with
\begin{align}
\varepsilon_\kk=\sqrt{A_\kk^2-B_\kk^2},
\end{align}
where
\begin{align}
A_\kk&=\frac{1}{2}(2\Delta J_\kk+\gamma_{\bq+\kk}+\gamma_{\bq-\kk}-4\gamma_\bq),\non\\
B_\kk&=\frac{1}{2}(\gamma_{\bq+\kk}+\gamma_{\bq-\kk}-2\Delta J_\kk).
\end{align}
The rest of the quadratic terms in $\mathcal{H}_{\mathrm{eff}}$ is obtained from a mean-field decoupling of the quartic Hamiltonian with
\begin{align}
\delta\varepsilon_\kk&=(u_\kk^2+v_\kk^2)\delta A_\kk+2u_\kk v_\kk\delta B_\kk,\non\\
O_\kk&=(u_\kk^2+v_\kk^2)\delta B_\kk+2u_\kk v_\kk\delta A_\kk,
\end{align}
where $u_\kk$ and $v_\kk$ are the Bogoliubov transformation coefficients given by
\begin{align}
u_{\kk}&=\sqrt{\frac{A_\kk}{2\varepsilon_{\kk}}+\frac{1}{2}},\non\\ v_{\kk}&=-\text{sgn}(B_\kk)\sqrt{\frac{A_\kk}{2\varepsilon_{\kk}}-\frac{1}{2}},
\end{align}
and
\begin{align}
  \delta A_\kk=&\frac{A_\kk}{2}+\frac{1}{2N}\sum_\pp \frac{1}{\varepsilon_\pp}\bigg[A_\pp \Big(A_{\kk-\pp}+B_{\kk-\pp}-A_\kk-A_\pp \Big)\non\\
  &+B_\pp \Big(\frac{B_\kk}{2}+B_\pp \Big) \bigg],\non\\
  \delta B_\kk=&\frac{B_\kk}{2}-\frac{1}{2N}\sum_\pp \frac{1}{\varepsilon_\pp}\bigg[B_\pp\Big(A_{\kk-\pp}+B_{\kk-\pp}-\frac{A_\kk}{2}-\frac{A_\pp}{2}\Big)\non\\
  &+A_\pp\Big(B_\kk+\frac{B_\pp}{2}\Big)\bigg].
\end{align}
The cubic interaction terms are defined as
\begin{align}\label{Eq:phia}
\Phi_{a}(1,2;3)=&\Big[\bar{\gamma}_1(u_1+v_1)(u_2u_3+v_2v_3)+\bar{\gamma}_2(u_2+v_2)(u_1u_3\non\\
  &+v_1v_3)-\bar{\gamma}_3(u_3+v_3)(u_1v_2+v_1u_2)\Big],\non\\
\Phi_{b}(1,2,3)=&\Big[\bar{\gamma}_1(u_1+v_1)(u_2v_3+v_2u_3)+\bar{\gamma}_2(u_2+v_2)(u_1v_3\non\\
  &+v_1u_3)+\bar{\gamma}_3(u_3+v_3)(u_1v_2+v_1u_2)\Big],
\end{align}
with
\begin{align}\label{Eq:gammabar}
\bar{\gamma}_\kk=\gamma_{\bq+\kk}-\gamma_{\bq-\kk}.
\end{align}
The quartic interaction term is given by
\begin{widetext}
\begin{eqnarray}
\Phi_{c}(1,2;3,4)=&&-(B_1+B_2+B_4)(u_1u_2u_3v_4+v_1v_2v_3u_4)
                 -(B_1+B_2+B_3)(u_1u_2v_3u_4+v_1v_2u_3v_4)\non\\
                 &&-(B_2+B_3+B_4)(u_1v_2u_3u_4+v_1u_2v_3v_4)
                 -(B_1+B_3+B_4)(u_1v_2v_3v_4+v_1u_2u_3u_4)\non\\
                 &&+[(C_{1-3}+C_{2-3}+C_{1-4}+C_{2-4})
                 -(A_1+A_2+A_3+A_4)](u_1u_2u_3u_4+v_1v_2v_3v_4)\non\\
                 &&+[(C_{1+2}+C_{3+4}+C_{1-3}+C_{2-4})
                 -(A_1+A_2+A_3+A_4)](u_1v_2u_3v_4+v_1u_2v_3u_4)\non\\
                 &&+[(C_{1+2}+C_{3+4}+C_{1-4}+C_{2-3})
                 -(A_1+A_2+A_3+A_4)](u_1v_2v_3u_4+v_1u_2u_3v_4),
\end{eqnarray}
\end{widetext}
where $C_\kk$ is
\begin{eqnarray}
C_\kk=A_\kk+B_\kk.
\end{eqnarray}

\bibliography{ref}

\begin{thebibliography}{65}%
\makeatletter
\providecommand \@ifxundefined [1]{%
 \@ifx{#1\undefined}
}%
\providecommand \@ifnum [1]{%
 \ifnum #1\expandafter \@firstoftwo
 \else \expandafter \@secondoftwo
 \fi
}%
\providecommand \@ifx [1]{%
 \ifx #1\expandafter \@firstoftwo
 \else \expandafter \@secondoftwo
 \fi
}%
\providecommand \natexlab [1]{#1}%
\providecommand \enquote  [1]{``#1''}%
\providecommand \bibnamefont  [1]{#1}%
\providecommand \bibfnamefont [1]{#1}%
\providecommand \citenamefont [1]{#1}%
\providecommand \href@noop [0]{\@secondoftwo}%
\providecommand \href [0]{\begingroup \@sanitize@url \@href}%
\providecommand \@href[1]{\@@startlink{#1}\@@href}%
\providecommand \@@href[1]{\endgroup#1\@@endlink}%
\providecommand \@sanitize@url [0]{\catcode `\\12\catcode `\$12\catcode
  `\&12\catcode `\#12\catcode `\^12\catcode `\_12\catcode `\%12\relax}%
\providecommand \@@startlink[1]{}%
\providecommand \@@endlink[0]{}%
\providecommand \url  [0]{\begingroup\@sanitize@url \@url }%
\providecommand \@url [1]{\endgroup\@href {#1}{\urlprefix }}%
\providecommand \urlprefix  [0]{URL }%
\providecommand \Eprint [0]{\href }%
\providecommand \doibase [0]{http://dx.doi.org/}%
\providecommand \selectlanguage [0]{\@gobble}%
\providecommand \bibinfo  [0]{\@secondoftwo}%
\providecommand \bibfield  [0]{\@secondoftwo}%
\providecommand \translation [1]{[#1]}%
\providecommand \BibitemOpen [0]{}%
\providecommand \bibitemStop [0]{}%
\providecommand \bibitemNoStop [0]{.\EOS\space}%
\providecommand \EOS [0]{\spacefactor3000\relax}%
\providecommand \BibitemShut  [1]{\csname bibitem#1\endcsname}%
\let\auto@bib@innerbib\@empty
\bibitem [{\citenamefont {Kohno}\ \emph {et~al.}(2007)\citenamefont {Kohno},
  \citenamefont {Starykh},\ and\ \citenamefont {Balents}}]{nphys749}%
  \BibitemOpen
  \bibfield  {author} {\bibinfo {author} {\bibfnamefont {Masanori}\
  \bibnamefont {Kohno}}, \bibinfo {author} {\bibfnamefont {Oleg~A.}\
  \bibnamefont {Starykh}}, \ and\ \bibinfo {author} {\bibfnamefont {Leon}\
  \bibnamefont {Balents}},\ }\bibfield  {title} {\enquote {\bibinfo {title}
  {Spinons and triplons in spatially anisotropic frustrated
  antiferromagnets},}\ }\href {\doibase 10.1038/nphys749} {\bibfield  {journal}
  {\bibinfo  {journal} {Nature Phys.}\ }\textbf {\bibinfo {volume} {3}},\
  \bibinfo {pages} {790} (\bibinfo {year} {2007})}\BibitemShut {NoStop}%
\bibitem [{\citenamefont {Swanson}\ \emph {et~al.}(2009)\citenamefont
  {Swanson}, \citenamefont {Haraldsen},\ and\ \citenamefont
  {Fishman}}]{PhysRevB.79.184413}%
  \BibitemOpen
  \bibfield  {author} {\bibinfo {author} {\bibfnamefont {M.}~\bibnamefont
  {Swanson}}, \bibinfo {author} {\bibfnamefont {J.~T.}\ \bibnamefont
  {Haraldsen}}, \ and\ \bibinfo {author} {\bibfnamefont {R.~S.}\ \bibnamefont
  {Fishman}},\ }\bibfield  {title} {\enquote {\bibinfo {title} {Critical
  anisotropies of a geometrically frustrated triangular-lattice
  antiferromagnet},}\ }\href {\doibase 10.1103/PhysRevB.79.184413} {\bibfield
  {journal} {\bibinfo  {journal} {Phys. Rev. B}\ }\textbf {\bibinfo {volume}
  {79}},\ \bibinfo {pages} {184413} (\bibinfo {year} {2009})}\BibitemShut
  {NoStop}%
\bibitem [{\citenamefont {Hauke}\ \emph {et~al.}(2011)\citenamefont {Hauke},
  \citenamefont {Roscilde}, \citenamefont {Murg}, \citenamefont {Cirac},\ and\
  \citenamefont {Schmied}}]{Hauke_2011}%
  \BibitemOpen
  \bibfield  {author} {\bibinfo {author} {\bibfnamefont {Philipp}\ \bibnamefont
  {Hauke}}, \bibinfo {author} {\bibfnamefont {Tommaso}\ \bibnamefont
  {Roscilde}}, \bibinfo {author} {\bibfnamefont {Valentin}\ \bibnamefont
  {Murg}}, \bibinfo {author} {\bibfnamefont {J~Ignacio}\ \bibnamefont {Cirac}},
  \ and\ \bibinfo {author} {\bibfnamefont {Roman}\ \bibnamefont {Schmied}},\
  }\bibfield  {title} {\enquote {\bibinfo {title} {Modified spin-wave theory
  with ordering vector optimization: spatially anisotropic triangular lattice
  and {J}$_1${J}$_2${J}$_3$ model with heisenberg interactions},}\ }\href
  {\doibase 10.1088/1367-2630/13/7/075017} {\bibfield  {journal} {\bibinfo
  {journal} {New Journal of Physics}\ }\textbf {\bibinfo {volume} {13}},\
  \bibinfo {pages} {075017} (\bibinfo {year} {2011})}\BibitemShut {NoStop}%
\bibitem [{\citenamefont {Weichselbaum}\ and\ \citenamefont
  {White}(2011)}]{PhysRevB.84.245130}%
  \BibitemOpen
  \bibfield  {author} {\bibinfo {author} {\bibfnamefont {Andreas}\ \bibnamefont
  {Weichselbaum}}\ and\ \bibinfo {author} {\bibfnamefont {Steven~R.}\
  \bibnamefont {White}},\ }\bibfield  {title} {\enquote {\bibinfo {title}
  {Incommensurate correlations in the anisotropic triangular heisenberg
  lattice},}\ }\href {\doibase 10.1103/PhysRevB.84.245130} {\bibfield
  {journal} {\bibinfo  {journal} {Phys. Rev. B}\ }\textbf {\bibinfo {volume}
  {84}},\ \bibinfo {pages} {245130} (\bibinfo {year} {2011})}\BibitemShut
  {NoStop}%
\bibitem [{\citenamefont {Chen}\ \emph {et~al.}(2013)\citenamefont {Chen},
  \citenamefont {Ju}, \citenamefont {Jiang}, \citenamefont {Starykh},\ and\
  \citenamefont {Balents}}]{PhysRevB.87.165123}%
  \BibitemOpen
  \bibfield  {author} {\bibinfo {author} {\bibfnamefont {Ru}~\bibnamefont
  {Chen}}, \bibinfo {author} {\bibfnamefont {Hyejin}\ \bibnamefont {Ju}},
  \bibinfo {author} {\bibfnamefont {Hong-Chen}\ \bibnamefont {Jiang}}, \bibinfo
  {author} {\bibfnamefont {Oleg~A.}\ \bibnamefont {Starykh}}, \ and\ \bibinfo
  {author} {\bibfnamefont {Leon}\ \bibnamefont {Balents}},\ }\bibfield  {title}
  {\enquote {\bibinfo {title} {Ground states of spin-$\frac{1}{2}$ triangular
  antiferromagnets in a magnetic field},}\ }\href {\doibase
  10.1103/PhysRevB.87.165123} {\bibfield  {journal} {\bibinfo  {journal} {Phys.
  Rev. B}\ }\textbf {\bibinfo {volume} {87}},\ \bibinfo {pages} {165123}
  (\bibinfo {year} {2013})}\BibitemShut {NoStop}%
\bibitem [{\citenamefont {Suzuki}\ \emph {et~al.}(2014)\citenamefont {Suzuki},
  \citenamefont {Matsubara}, \citenamefont {Fujiki},\ and\ \citenamefont
  {Shirakura}}]{PhysRevB.90.184414}%
  \BibitemOpen
  \bibfield  {author} {\bibinfo {author} {\bibfnamefont {Nobuo}\ \bibnamefont
  {Suzuki}}, \bibinfo {author} {\bibfnamefont {Fumitaka}\ \bibnamefont
  {Matsubara}}, \bibinfo {author} {\bibfnamefont {Sumiyoshi}\ \bibnamefont
  {Fujiki}}, \ and\ \bibinfo {author} {\bibfnamefont {Takayuki}\ \bibnamefont
  {Shirakura}},\ }\bibfield  {title} {\enquote {\bibinfo {title} {Absence of
  classical long-range order in an $s=\frac{1}{2}$ heisenberg antiferromagnet
  on a triangular lattice},}\ }\href {\doibase 10.1103/PhysRevB.90.184414}
  {\bibfield  {journal} {\bibinfo  {journal} {Phys. Rev. B}\ }\textbf {\bibinfo
  {volume} {90}},\ \bibinfo {pages} {184414} (\bibinfo {year}
  {2014})}\BibitemShut {NoStop}%
\bibitem [{\citenamefont {Schmidt}\ and\ \citenamefont
  {Thalmeier}(2014)}]{SchmidtPhysRevB.89.184402}%
  \BibitemOpen
  \bibfield  {author} {\bibinfo {author} {\bibfnamefont {Burkhard}\
  \bibnamefont {Schmidt}}\ and\ \bibinfo {author} {\bibfnamefont {Peter}\
  \bibnamefont {Thalmeier}},\ }\bibfield  {title} {\enquote {\bibinfo {title}
  {Quantum fluctuations in anisotropic triangular lattices with ferromagnetic
  and antiferromagnetic exchange},}\ }\href {\doibase
  10.1103/PhysRevB.89.184402} {\bibfield  {journal} {\bibinfo  {journal} {Phys.
  Rev. B}\ }\textbf {\bibinfo {volume} {89}},\ \bibinfo {pages} {184402}
  (\bibinfo {year} {2014})}\BibitemShut {NoStop}%
\bibitem [{\citenamefont {Starykh}\ \emph {et~al.}(2014)\citenamefont
  {Starykh}, \citenamefont {Jin},\ and\ \citenamefont
  {Chubukov}}]{StarykhPhysRevLett.113.087204}%
  \BibitemOpen
  \bibfield  {author} {\bibinfo {author} {\bibfnamefont {Oleg~A.}\ \bibnamefont
  {Starykh}}, \bibinfo {author} {\bibfnamefont {Wen}\ \bibnamefont {Jin}}, \
  and\ \bibinfo {author} {\bibfnamefont {Andrey~V.}\ \bibnamefont {Chubukov}},\
  }\bibfield  {title} {\enquote {\bibinfo {title} {Phases of a
  triangular-lattice antiferromagnet near saturation},}\ }\href {\doibase
  10.1103/PhysRevLett.113.087204} {\bibfield  {journal} {\bibinfo  {journal}
  {Phys. Rev. Lett.}\ }\textbf {\bibinfo {volume} {113}},\ \bibinfo {pages}
  {087204} (\bibinfo {year} {2014})}\BibitemShut {NoStop}%
\bibitem [{\citenamefont {Balents}(2010)}]{nature08917}%
  \BibitemOpen
  \bibfield  {author} {\bibinfo {author} {\bibfnamefont {Leon}\ \bibnamefont
  {Balents}},\ }\bibfield  {title} {\enquote {\bibinfo {title} {Spin liquids in
  frustrated magnets},}\ }\href {\doibase 10.1038/nature08917} {\bibfield
  {journal} {\bibinfo  {journal} {Nature}\ }\textbf {\bibinfo {volume} {464}},\
  \bibinfo {pages} {199} (\bibinfo {year} {2010})}\BibitemShut {NoStop}%
\bibitem [{\citenamefont {Schollwck}\ \emph {et~al.}(2004)\citenamefont
  {Schollwck}, \citenamefont {Richter}, \citenamefont {Farnell},\ and\
  \citenamefont {Bishop}}]{b96825}%
  \BibitemOpen
  \bibfield  {author} {\bibinfo {author} {\bibfnamefont {Ulrich}\ \bibnamefont
  {Schollwck}}, \bibinfo {author} {\bibfnamefont {Johannes}\ \bibnamefont
  {Richter}}, \bibinfo {author} {\bibfnamefont {Damian J.~J.}\ \bibnamefont
  {Farnell}}, \ and\ \bibinfo {author} {\bibfnamefont {Raymod~F.}\ \bibnamefont
  {Bishop}},\ }\bibfield  {title} {\enquote {\bibinfo {title} {Quantum
  magnetism},}\ }\href {\doibase 10.1007/b96825} {\bibfield  {journal}
  {\bibinfo  {journal} {Lecture Notes in Physics}\ }\textbf {\bibinfo {volume}
  {645}} (\bibinfo {year} {2004}),\ 10.1007/b96825}\BibitemShut {NoStop}%
\bibitem [{\citenamefont {Kanoda}\ and\ \citenamefont
  {Kato}(2011)}]{doi:10.1146/annurev-conmatphys-062910-140521}%
  \BibitemOpen
  \bibfield  {author} {\bibinfo {author} {\bibfnamefont {Kazushi}\ \bibnamefont
  {Kanoda}}\ and\ \bibinfo {author} {\bibfnamefont {Reizo}\ \bibnamefont
  {Kato}},\ }\bibfield  {title} {\enquote {\bibinfo {title} {Mott physics in
  organic conductors with triangular lattices},}\ }\href {\doibase
  10.1146/annurev-conmatphys-062910-140521} {\bibfield  {journal} {\bibinfo
  {journal} {Annual Review of Condensed Matter Physics}\ }\textbf {\bibinfo
  {volume} {2}},\ \bibinfo {pages} {167--188} (\bibinfo {year}
  {2011})}\BibitemShut {NoStop}%
\bibitem [{\citenamefont {Powell}\ and\ \citenamefont
  {McKenzie}(2011)}]{Powell_2011}%
  \BibitemOpen
  \bibfield  {author} {\bibinfo {author} {\bibfnamefont {B~J}\ \bibnamefont
  {Powell}}\ and\ \bibinfo {author} {\bibfnamefont {Ross~H}\ \bibnamefont
  {McKenzie}},\ }\bibfield  {title} {\enquote {\bibinfo {title} {Quantum
  frustration in organic mott insulators: from spin liquids to unconventional
  superconductors},}\ }\href {\doibase 10.1088/0034-4885/74/5/056501}
  {\bibfield  {journal} {\bibinfo  {journal} {Reports on Progress in Physics}\
  }\textbf {\bibinfo {volume} {74}},\ \bibinfo {pages} {056501} (\bibinfo
  {year} {2011})}\BibitemShut {NoStop}%
\bibitem [{\citenamefont {Du}\ \emph {et~al.}(2015)\citenamefont {Du},
  \citenamefont {Liu}, \citenamefont {Xie}, \citenamefont {Wang},\ and\
  \citenamefont {Liu}}]{PhysRevB.92.214409}%
  \BibitemOpen
  \bibfield  {author} {\bibinfo {author} {\bibfnamefont {Z.~Z.}\ \bibnamefont
  {Du}}, \bibinfo {author} {\bibfnamefont {H.~M.}\ \bibnamefont {Liu}},
  \bibinfo {author} {\bibfnamefont {Y.~L.}\ \bibnamefont {Xie}}, \bibinfo
  {author} {\bibfnamefont {Q.~H.}\ \bibnamefont {Wang}}, \ and\ \bibinfo
  {author} {\bibfnamefont {J.-M.}\ \bibnamefont {Liu}},\ }\bibfield  {title}
  {\enquote {\bibinfo {title} {Spin casimir effect in noncollinear quantum
  antiferromagnets: Torque equilibrium spin wave approach},}\ }\href {\doibase
  10.1103/PhysRevB.92.214409} {\bibfield  {journal} {\bibinfo  {journal} {Phys.
  Rev. B}\ }\textbf {\bibinfo {volume} {92}},\ \bibinfo {pages} {214409}
  (\bibinfo {year} {2015})}\BibitemShut {NoStop}%
\bibitem [{\citenamefont {Du}\ \emph {et~al.}(2016)\citenamefont {Du},
  \citenamefont {Liu}, \citenamefont {Xie}, \citenamefont {Wang},\ and\
  \citenamefont {Liu}}]{PhysRevB.94.134416}%
  \BibitemOpen
  \bibfield  {author} {\bibinfo {author} {\bibfnamefont {Z.~Z.}\ \bibnamefont
  {Du}}, \bibinfo {author} {\bibfnamefont {H.~M.}\ \bibnamefont {Liu}},
  \bibinfo {author} {\bibfnamefont {Y.~L.}\ \bibnamefont {Xie}}, \bibinfo
  {author} {\bibfnamefont {Q.~H.}\ \bibnamefont {Wang}}, \ and\ \bibinfo
  {author} {\bibfnamefont {J.-M.}\ \bibnamefont {Liu}},\ }\bibfield  {title}
  {\enquote {\bibinfo {title} {Magnetic excitations in quasi-one-dimensional
  helimagnets: Magnon decays and influence of interchain interactions},}\
  }\href {\doibase 10.1103/PhysRevB.94.134416} {\bibfield  {journal} {\bibinfo
  {journal} {Phys. Rev. B}\ }\textbf {\bibinfo {volume} {94}},\ \bibinfo
  {pages} {134416} (\bibinfo {year} {2016})}\BibitemShut {NoStop}%
\bibitem [{\citenamefont {Ma}\ \emph {et~al.}(2016)\citenamefont {Ma},
  \citenamefont {Kamiya}, \citenamefont {Hong}, \citenamefont {Cao},
  \citenamefont {Ehlers}, \citenamefont {Tian}, \citenamefont {Batista},
  \citenamefont {Dun}, \citenamefont {Zhou},\ and\ \citenamefont
  {Matsuda}}]{PhysRevLett.116.087201}%
  \BibitemOpen
  \bibfield  {author} {\bibinfo {author} {\bibfnamefont {J.}~\bibnamefont
  {Ma}}, \bibinfo {author} {\bibfnamefont {Y.}~\bibnamefont {Kamiya}}, \bibinfo
  {author} {\bibfnamefont {Tao}\ \bibnamefont {Hong}}, \bibinfo {author}
  {\bibfnamefont {H.~B.}\ \bibnamefont {Cao}}, \bibinfo {author} {\bibfnamefont
  {G.}~\bibnamefont {Ehlers}}, \bibinfo {author} {\bibfnamefont
  {W.}~\bibnamefont {Tian}}, \bibinfo {author} {\bibfnamefont {C.~D.}\
  \bibnamefont {Batista}}, \bibinfo {author} {\bibfnamefont {Z.~L.}\
  \bibnamefont {Dun}}, \bibinfo {author} {\bibfnamefont {H.~D.}\ \bibnamefont
  {Zhou}}, \ and\ \bibinfo {author} {\bibfnamefont {M.}~\bibnamefont
  {Matsuda}},\ }\bibfield  {title} {\enquote {\bibinfo {title} {Static and
  dynamical properties of the spin-$1/2$ equilateral triangular-lattice
  antiferromagnet {Ba}$_{3}${CoSb}$_{2}${O}$_{9}$},}\ }\href {\doibase
  10.1103/PhysRevLett.116.087201} {\bibfield  {journal} {\bibinfo  {journal}
  {Phys. Rev. Lett.}\ }\textbf {\bibinfo {volume} {116}},\ \bibinfo {pages}
  {087201} (\bibinfo {year} {2016})}\BibitemShut {NoStop}%
\bibitem [{\citenamefont {Toth}\ \emph {et~al.}(2012)\citenamefont {Toth},
  \citenamefont {Lake}, \citenamefont {Hradil}, \citenamefont {Guidi},
  \citenamefont {Rule}, \citenamefont {Stone},\ and\ \citenamefont
  {Islam}}]{PhysRevLett.109.127203}%
  \BibitemOpen
  \bibfield  {author} {\bibinfo {author} {\bibfnamefont {S.}~\bibnamefont
  {Toth}}, \bibinfo {author} {\bibfnamefont {B.}~\bibnamefont {Lake}}, \bibinfo
  {author} {\bibfnamefont {K.}~\bibnamefont {Hradil}}, \bibinfo {author}
  {\bibfnamefont {T.}~\bibnamefont {Guidi}}, \bibinfo {author} {\bibfnamefont
  {K.~C.}\ \bibnamefont {Rule}}, \bibinfo {author} {\bibfnamefont {M.~B.}\
  \bibnamefont {Stone}}, \ and\ \bibinfo {author} {\bibfnamefont {A.~T. M.~N.}\
  \bibnamefont {Islam}},\ }\bibfield  {title} {\enquote {\bibinfo {title}
  {Magnetic soft modes in the distorted triangular antiferromagnet
  $\alpha$-{GaCr}$_{2}${O}$_{4}$},}\ }\href {\doibase
  10.1103/PhysRevLett.109.127203} {\bibfield  {journal} {\bibinfo  {journal}
  {Phys. Rev. Lett.}\ }\textbf {\bibinfo {volume} {109}},\ \bibinfo {pages}
  {127203} (\bibinfo {year} {2012})}\BibitemShut {NoStop}%
\bibitem [{\citenamefont {Songvilay}\ \emph {et~al.}(2017)\citenamefont
  {Songvilay}, \citenamefont {Petit}, \citenamefont {Suard}, \citenamefont
  {Martin},\ and\ \citenamefont {Damay}}]{PhysRevB.96.024416}%
  \BibitemOpen
  \bibfield  {author} {\bibinfo {author} {\bibfnamefont {M.}~\bibnamefont
  {Songvilay}}, \bibinfo {author} {\bibfnamefont {S.}~\bibnamefont {Petit}},
  \bibinfo {author} {\bibfnamefont {E.}~\bibnamefont {Suard}}, \bibinfo
  {author} {\bibfnamefont {C.}~\bibnamefont {Martin}}, \ and\ \bibinfo {author}
  {\bibfnamefont {F.}~\bibnamefont {Damay}},\ }\bibfield  {title} {\enquote
  {\bibinfo {title} {Spin dynamics in the distorted triangular lattice
  antiferromagnet $\alpha$-{SrCr}$_{2}${O}$_{4}$},}\ }\href {\doibase
  10.1103/PhysRevB.96.024416} {\bibfield  {journal} {\bibinfo  {journal} {Phys.
  Rev. B}\ }\textbf {\bibinfo {volume} {96}},\ \bibinfo {pages} {024416}
  (\bibinfo {year} {2017})}\BibitemShut {NoStop}%
\bibitem [{\citenamefont {Zheng}\ \emph
  {et~al.}(2006{\natexlab{a}})\citenamefont {Zheng}, \citenamefont
  {Fj\ae{}restad}, \citenamefont {Singh}, \citenamefont {McKenzie},\ and\
  \citenamefont {Coldea}}]{PhysRevLett.96.057201}%
  \BibitemOpen
  \bibfield  {author} {\bibinfo {author} {\bibfnamefont {Weihong}\ \bibnamefont
  {Zheng}}, \bibinfo {author} {\bibfnamefont {John~O.}\ \bibnamefont
  {Fj\ae{}restad}}, \bibinfo {author} {\bibfnamefont {Rajiv R.~P.}\
  \bibnamefont {Singh}}, \bibinfo {author} {\bibfnamefont {Ross~H.}\
  \bibnamefont {McKenzie}}, \ and\ \bibinfo {author} {\bibfnamefont {Radu}\
  \bibnamefont {Coldea}},\ }\bibfield  {title} {\enquote {\bibinfo {title}
  {Anomalous excitation spectra of frustrated quantum antiferromagnets},}\
  }\href {\doibase 10.1103/PhysRevLett.96.057201} {\bibfield  {journal}
  {\bibinfo  {journal} {Phys. Rev. Lett.}\ }\textbf {\bibinfo {volume} {96}},\
  \bibinfo {pages} {057201} (\bibinfo {year} {2006}{\natexlab{a}})}\BibitemShut
  {NoStop}%
\bibitem [{\citenamefont {Zheng}\ \emph
  {et~al.}(2006{\natexlab{b}})\citenamefont {Zheng}, \citenamefont
  {Fj\ae{}restad}, \citenamefont {Singh}, \citenamefont {McKenzie},\ and\
  \citenamefont {Coldea}}]{PhysRevB.74.224420}%
  \BibitemOpen
  \bibfield  {author} {\bibinfo {author} {\bibfnamefont {Weihong}\ \bibnamefont
  {Zheng}}, \bibinfo {author} {\bibfnamefont {John~O.}\ \bibnamefont
  {Fj\ae{}restad}}, \bibinfo {author} {\bibfnamefont {Rajiv R.~P.}\
  \bibnamefont {Singh}}, \bibinfo {author} {\bibfnamefont {Ross~H.}\
  \bibnamefont {McKenzie}}, \ and\ \bibinfo {author} {\bibfnamefont {Radu}\
  \bibnamefont {Coldea}},\ }\bibfield  {title} {\enquote {\bibinfo {title}
  {Excitation spectra of the spin-$\frac{1}{2}$ triangular-lattice heisenberg
  antiferromagnet},}\ }\href {\doibase 10.1103/PhysRevB.74.224420} {\bibfield
  {journal} {\bibinfo  {journal} {Phys. Rev. B}\ }\textbf {\bibinfo {volume}
  {74}},\ \bibinfo {pages} {224420} (\bibinfo {year}
  {2006}{\natexlab{b}})}\BibitemShut {NoStop}%
\bibitem [{\citenamefont {Coldea}\ \emph {et~al.}(2003)\citenamefont {Coldea},
  \citenamefont {Tennant},\ and\ \citenamefont
  {Tylczynski}}]{PhysRevB.68.134424}%
  \BibitemOpen
  \bibfield  {author} {\bibinfo {author} {\bibfnamefont {R.}~\bibnamefont
  {Coldea}}, \bibinfo {author} {\bibfnamefont {D.~A.}\ \bibnamefont {Tennant}},
  \ and\ \bibinfo {author} {\bibfnamefont {Z.}~\bibnamefont {Tylczynski}},\
  }\bibfield  {title} {\enquote {\bibinfo {title} {Extended scattering continua
  characteristic of spin fractionalization in the two-dimensional frustrated
  quantum magnet {Cs}$_{2}${CuCl}$_{4}$ observed by neutron scattering},}\
  }\href {\doibase 10.1103/PhysRevB.68.134424} {\bibfield  {journal} {\bibinfo
  {journal} {Phys. Rev. B}\ }\textbf {\bibinfo {volume} {68}},\ \bibinfo
  {pages} {134424} (\bibinfo {year} {2003})}\BibitemShut {NoStop}%
\bibitem [{\citenamefont {Zhou}\ \emph {et~al.}(2012)\citenamefont {Zhou},
  \citenamefont {Xu}, \citenamefont {Hallas}, \citenamefont {Silverstein},
  \citenamefont {Wiebe}, \citenamefont {Umegaki}, \citenamefont {Yan},
  \citenamefont {Murphy}, \citenamefont {Park}, \citenamefont {Qiu},
  \citenamefont {Copley}, \citenamefont {Gardner},\ and\ \citenamefont
  {Takano}}]{PhysRevLett.109.267206}%
  \BibitemOpen
  \bibfield  {author} {\bibinfo {author} {\bibfnamefont {H.~D.}\ \bibnamefont
  {Zhou}}, \bibinfo {author} {\bibfnamefont {Cenke}\ \bibnamefont {Xu}},
  \bibinfo {author} {\bibfnamefont {A.~M.}\ \bibnamefont {Hallas}}, \bibinfo
  {author} {\bibfnamefont {H.~J.}\ \bibnamefont {Silverstein}}, \bibinfo
  {author} {\bibfnamefont {C.~R.}\ \bibnamefont {Wiebe}}, \bibinfo {author}
  {\bibfnamefont {I.}~\bibnamefont {Umegaki}}, \bibinfo {author} {\bibfnamefont
  {J.~Q.}\ \bibnamefont {Yan}}, \bibinfo {author} {\bibfnamefont {T.~P.}\
  \bibnamefont {Murphy}}, \bibinfo {author} {\bibfnamefont {J.-H.}\
  \bibnamefont {Park}}, \bibinfo {author} {\bibfnamefont {Y.}~\bibnamefont
  {Qiu}}, \bibinfo {author} {\bibfnamefont {J.~R.~D.}\ \bibnamefont {Copley}},
  \bibinfo {author} {\bibfnamefont {J.~S.}\ \bibnamefont {Gardner}}, \ and\
  \bibinfo {author} {\bibfnamefont {Y.}~\bibnamefont {Takano}},\ }\bibfield
  {title} {\enquote {\bibinfo {title} {Successive phase transitions and
  extended spin-excitation continuum in the $s\mathbf{=}\frac{1}{2}$
  triangular-lattice antiferromagnet {Ba}$_3${CoSb}$_2${O}$_9$},}\ }\href
  {\doibase 10.1103/PhysRevLett.109.267206} {\bibfield  {journal} {\bibinfo
  {journal} {Phys. Rev. Lett.}\ }\textbf {\bibinfo {volume} {109}},\ \bibinfo
  {pages} {267206} (\bibinfo {year} {2012})}\BibitemShut {NoStop}%
\bibitem [{\citenamefont {Oh}\ \emph {et~al.}(2013)\citenamefont {Oh},
  \citenamefont {Le}, \citenamefont {Jeong}, \citenamefont {Lee}, \citenamefont
  {Woo}, \citenamefont {Song}, \citenamefont {Perring}, \citenamefont {Buyers},
  \citenamefont {Cheong},\ and\ \citenamefont {Park}}]{PhysRevLett.111.257202}%
  \BibitemOpen
  \bibfield  {author} {\bibinfo {author} {\bibfnamefont {Joosung}\ \bibnamefont
  {Oh}}, \bibinfo {author} {\bibfnamefont {Manh~Duc}\ \bibnamefont {Le}},
  \bibinfo {author} {\bibfnamefont {Jaehong}\ \bibnamefont {Jeong}}, \bibinfo
  {author} {\bibfnamefont {Jung-hyun}\ \bibnamefont {Lee}}, \bibinfo {author}
  {\bibfnamefont {Hyungje}\ \bibnamefont {Woo}}, \bibinfo {author}
  {\bibfnamefont {Wan-Young}\ \bibnamefont {Song}}, \bibinfo {author}
  {\bibfnamefont {T.~G.}\ \bibnamefont {Perring}}, \bibinfo {author}
  {\bibfnamefont {W.~J.~L.}\ \bibnamefont {Buyers}}, \bibinfo {author}
  {\bibfnamefont {S.-W.}\ \bibnamefont {Cheong}}, \ and\ \bibinfo {author}
  {\bibfnamefont {Je-Geun}\ \bibnamefont {Park}},\ }\bibfield  {title}
  {\enquote {\bibinfo {title} {Magnon breakdown in a two dimensional triangular
  lattice heisenberg antiferromagnet of multiferroic {LuMnO}$_3$},}\ }\href
  {\doibase 10.1103/PhysRevLett.111.257202} {\bibfield  {journal} {\bibinfo
  {journal} {Phys. Rev. Lett.}\ }\textbf {\bibinfo {volume} {111}},\ \bibinfo
  {pages} {257202} (\bibinfo {year} {2013})}\BibitemShut {NoStop}%
\bibitem [{\citenamefont {Feynman}(1998)}]{book:780611}%
  \BibitemOpen
  \bibfield  {author} {\bibinfo {author} {\bibfnamefont {Richard~P.}\
  \bibnamefont {Feynman}},\ }\href@noop {} {\emph {\bibinfo {title}
  {Statistical Mechanics: A Set Of Lectures (Advanced Books Classics)}}},\
  \bibinfo {edition} {2nd}\ ed.,\ Advanced Books Classics\ (\bibinfo
  {publisher} {Westview Press},\ \bibinfo {year} {1998})\BibitemShut {NoStop}%
\bibitem [{\citenamefont {Girvin}\ \emph {et~al.}(1986)\citenamefont {Girvin},
  \citenamefont {MacDonald},\ and\ \citenamefont
  {Platzman}}]{PhysRevB.33.2481}%
  \BibitemOpen
  \bibfield  {author} {\bibinfo {author} {\bibfnamefont {S.~M.}\ \bibnamefont
  {Girvin}}, \bibinfo {author} {\bibfnamefont {A.~H.}\ \bibnamefont
  {MacDonald}}, \ and\ \bibinfo {author} {\bibfnamefont {P.~M.}\ \bibnamefont
  {Platzman}},\ }\bibfield  {title} {\enquote {\bibinfo {title} {Magneto-roton
  theory of collective excitations in the fractional quantum hall effect},}\
  }\href {\doibase 10.1103/PhysRevB.33.2481} {\bibfield  {journal} {\bibinfo
  {journal} {Phys. Rev. B}\ }\textbf {\bibinfo {volume} {33}},\ \bibinfo
  {pages} {2481--2494} (\bibinfo {year} {1986})}\BibitemShut {NoStop}%
\bibitem [{\citenamefont {Kubo}\ and\ \citenamefont
  {Kurihara}(2014)}]{PhysRevB.90.014421}%
  \BibitemOpen
  \bibfield  {author} {\bibinfo {author} {\bibfnamefont {Yurika}\ \bibnamefont
  {Kubo}}\ and\ \bibinfo {author} {\bibfnamefont {Susumu}\ \bibnamefont
  {Kurihara}},\ }\bibfield  {title} {\enquote {\bibinfo {title} {Tunable rotons
  in square-lattice antiferromagnets under strong magnetic fields},}\ }\href
  {\doibase 10.1103/PhysRevB.90.014421} {\bibfield  {journal} {\bibinfo
  {journal} {Phys. Rev. B}\ }\textbf {\bibinfo {volume} {90}},\ \bibinfo
  {pages} {014421} (\bibinfo {year} {2014})}\BibitemShut {NoStop}%
\bibitem [{\citenamefont {Powalski}\ \emph {et~al.}(2015)\citenamefont
  {Powalski}, \citenamefont {Uhrig},\ and\ \citenamefont
  {Schmidt}}]{PhysRevLett.115.207202}%
  \BibitemOpen
  \bibfield  {author} {\bibinfo {author} {\bibfnamefont {M.}~\bibnamefont
  {Powalski}}, \bibinfo {author} {\bibfnamefont {G.~S.}\ \bibnamefont {Uhrig}},
  \ and\ \bibinfo {author} {\bibfnamefont {K.~P.}\ \bibnamefont {Schmidt}},\
  }\bibfield  {title} {\enquote {\bibinfo {title} {Roton minimum as a
  fingerprint of magnon-higgs scattering in ordered quantum
  antiferromagnets},}\ }\href {\doibase 10.1103/PhysRevLett.115.207202}
  {\bibfield  {journal} {\bibinfo  {journal} {Phys. Rev. Lett.}\ }\textbf
  {\bibinfo {volume} {115}},\ \bibinfo {pages} {207202} (\bibinfo {year}
  {2015})}\BibitemShut {NoStop}%
\bibitem [{\citenamefont {Starykh}\ \emph {et~al.}(2006)\citenamefont
  {Starykh}, \citenamefont {Chubukov},\ and\ \citenamefont
  {Abanov}}]{PhysRevB.74.180403}%
  \BibitemOpen
  \bibfield  {author} {\bibinfo {author} {\bibfnamefont {Oleg~A.}\ \bibnamefont
  {Starykh}}, \bibinfo {author} {\bibfnamefont {Andrey~V.}\ \bibnamefont
  {Chubukov}}, \ and\ \bibinfo {author} {\bibfnamefont {Alexander~G.}\
  \bibnamefont {Abanov}},\ }\bibfield  {title} {\enquote {\bibinfo {title}
  {Flat spin-wave dispersion in a triangular antiferromagnet},}\ }\href
  {\doibase 10.1103/PhysRevB.74.180403} {\bibfield  {journal} {\bibinfo
  {journal} {Phys. Rev. B}\ }\textbf {\bibinfo {volume} {74}},\ \bibinfo
  {pages} {180403} (\bibinfo {year} {2006})}\BibitemShut {NoStop}%
\bibitem [{\citenamefont {Chernyshev}\ and\ \citenamefont
  {Zhitomirsky}(2009)}]{PhysRevB.79.144416}%
  \BibitemOpen
  \bibfield  {author} {\bibinfo {author} {\bibfnamefont {A.~L.}\ \bibnamefont
  {Chernyshev}}\ and\ \bibinfo {author} {\bibfnamefont {M.~E.}\ \bibnamefont
  {Zhitomirsky}},\ }\bibfield  {title} {\enquote {\bibinfo {title} {Spin waves
  in a triangular lattice antiferromagnet: Decays, spectrum renormalization,
  and singularities},}\ }\href {\doibase 10.1103/PhysRevB.79.144416} {\bibfield
   {journal} {\bibinfo  {journal} {Phys. Rev. B}\ }\textbf {\bibinfo {volume}
  {79}},\ \bibinfo {pages} {144416} (\bibinfo {year} {2009})}\BibitemShut
  {NoStop}%
\bibitem [{\citenamefont {Isakov}\ \emph {et~al.}(2005)\citenamefont {Isakov},
  \citenamefont {Senthil},\ and\ \citenamefont {Kim}}]{PhysRevB.72.174417}%
  \BibitemOpen
  \bibfield  {author} {\bibinfo {author} {\bibfnamefont {S.~V.}\ \bibnamefont
  {Isakov}}, \bibinfo {author} {\bibfnamefont {T.}~\bibnamefont {Senthil}}, \
  and\ \bibinfo {author} {\bibfnamefont {Yong~Baek}\ \bibnamefont {Kim}},\
  }\bibfield  {title} {\enquote {\bibinfo {title} {Ordering in
  {Cs}$_{2}${CuCl}$_{4}$: Possibility of a proximate spin liquid},}\ }\href
  {\doibase 10.1103/PhysRevB.72.174417} {\bibfield  {journal} {\bibinfo
  {journal} {Phys. Rev. B}\ }\textbf {\bibinfo {volume} {72}},\ \bibinfo
  {pages} {174417} (\bibinfo {year} {2005})}\BibitemShut {NoStop}%
\bibitem [{\citenamefont {Alicea}\ \emph {et~al.}(2005)\citenamefont {Alicea},
  \citenamefont {Motrunich},\ and\ \citenamefont
  {Fisher}}]{PhysRevLett.95.247203}%
  \BibitemOpen
  \bibfield  {author} {\bibinfo {author} {\bibfnamefont {Jason}\ \bibnamefont
  {Alicea}}, \bibinfo {author} {\bibfnamefont {Olexei~I.}\ \bibnamefont
  {Motrunich}}, \ and\ \bibinfo {author} {\bibfnamefont {Matthew P.~A.}\
  \bibnamefont {Fisher}},\ }\bibfield  {title} {\enquote {\bibinfo {title}
  {Algebraic vortex liquid in spin-$1/2$ triangular antiferromagnets: Scenario
  for {Cs}$_{2}${CuCl}$_{4}$},}\ }\href {\doibase
  10.1103/PhysRevLett.95.247203} {\bibfield  {journal} {\bibinfo  {journal}
  {Phys. Rev. Lett.}\ }\textbf {\bibinfo {volume} {95}},\ \bibinfo {pages}
  {247203} (\bibinfo {year} {2005})}\BibitemShut {NoStop}%
\bibitem [{\citenamefont {Yunoki}\ and\ \citenamefont
  {Sorella}(2006)}]{PhysRevB.74.014408}%
  \BibitemOpen
  \bibfield  {author} {\bibinfo {author} {\bibfnamefont {Seiji}\ \bibnamefont
  {Yunoki}}\ and\ \bibinfo {author} {\bibfnamefont {Sandro}\ \bibnamefont
  {Sorella}},\ }\bibfield  {title} {\enquote {\bibinfo {title} {Two spin liquid
  phases in the spatially anisotropic triangular heisenberg model},}\ }\href
  {\doibase 10.1103/PhysRevB.74.014408} {\bibfield  {journal} {\bibinfo
  {journal} {Phys. Rev. B}\ }\textbf {\bibinfo {volume} {74}},\ \bibinfo
  {pages} {014408} (\bibinfo {year} {2006})}\BibitemShut {NoStop}%
\bibitem [{\citenamefont {Starykh}\ and\ \citenamefont
  {Balents}(2007)}]{PhysRevLett.98.077205}%
  \BibitemOpen
  \bibfield  {author} {\bibinfo {author} {\bibfnamefont {Oleg~A.}\ \bibnamefont
  {Starykh}}\ and\ \bibinfo {author} {\bibfnamefont {Leon}\ \bibnamefont
  {Balents}},\ }\bibfield  {title} {\enquote {\bibinfo {title} {Ordering in
  spatially anisotropic triangular antiferromagnets},}\ }\href {\doibase
  10.1103/PhysRevLett.98.077205} {\bibfield  {journal} {\bibinfo  {journal}
  {Phys. Rev. Lett.}\ }\textbf {\bibinfo {volume} {98}},\ \bibinfo {pages}
  {077205} (\bibinfo {year} {2007})}\BibitemShut {NoStop}%
\bibitem [{\citenamefont {Ghioldi}\ \emph {et~al.}(2015)\citenamefont
  {Ghioldi}, \citenamefont {Mezio}, \citenamefont {Manuel}, \citenamefont
  {Singh}, \citenamefont {Oitmaa},\ and\ \citenamefont
  {Trumper}}]{PhysRevB.91.134423}%
  \BibitemOpen
  \bibfield  {author} {\bibinfo {author} {\bibfnamefont {E.~A.}\ \bibnamefont
  {Ghioldi}}, \bibinfo {author} {\bibfnamefont {A.}~\bibnamefont {Mezio}},
  \bibinfo {author} {\bibfnamefont {L.~O.}\ \bibnamefont {Manuel}}, \bibinfo
  {author} {\bibfnamefont {R.~R.~P.}\ \bibnamefont {Singh}}, \bibinfo {author}
  {\bibfnamefont {J.}~\bibnamefont {Oitmaa}}, \ and\ \bibinfo {author}
  {\bibfnamefont {A.~E.}\ \bibnamefont {Trumper}},\ }\bibfield  {title}
  {\enquote {\bibinfo {title} {Magnons and excitation continuum in xxz
  triangular antiferromagnetic model: Application to
  {Ba}$_3${CoSb}$_2${O}$_9$},}\ }\href {\doibase 10.1103/PhysRevB.91.134423}
  {\bibfield  {journal} {\bibinfo  {journal} {Phys. Rev. B}\ }\textbf {\bibinfo
  {volume} {91}},\ \bibinfo {pages} {134423} (\bibinfo {year}
  {2015})}\BibitemShut {NoStop}%
\bibitem [{\citenamefont {Veillette}\ \emph {et~al.}(2005)\citenamefont
  {Veillette}, \citenamefont {James},\ and\ \citenamefont
  {Essler}}]{PhysRevB.72.134429}%
  \BibitemOpen
  \bibfield  {author} {\bibinfo {author} {\bibfnamefont {M.~Y.}\ \bibnamefont
  {Veillette}}, \bibinfo {author} {\bibfnamefont {A.~J.~A.}\ \bibnamefont
  {James}}, \ and\ \bibinfo {author} {\bibfnamefont {F.~H.~L.}\ \bibnamefont
  {Essler}},\ }\bibfield  {title} {\enquote {\bibinfo {title} {Spin dynamics of
  the quasi-two-dimensional spin-$\frac{1}{2}$ quantum magnet
  {Cs}$_{2}${CuCl}$_{4}$},}\ }\href {\doibase 10.1103/PhysRevB.72.134429}
  {\bibfield  {journal} {\bibinfo  {journal} {Phys. Rev. B}\ }\textbf {\bibinfo
  {volume} {72}},\ \bibinfo {pages} {134429} (\bibinfo {year}
  {2005})}\BibitemShut {NoStop}%
\bibitem [{\citenamefont {Dalidovich}\ \emph {et~al.}(2006)\citenamefont
  {Dalidovich}, \citenamefont {Sknepnek}, \citenamefont {Berlinsky},
  \citenamefont {Zhang},\ and\ \citenamefont {Kallin}}]{PhysRevB.73.184403}%
  \BibitemOpen
  \bibfield  {author} {\bibinfo {author} {\bibfnamefont {Denis}\ \bibnamefont
  {Dalidovich}}, \bibinfo {author} {\bibfnamefont {Rastko}\ \bibnamefont
  {Sknepnek}}, \bibinfo {author} {\bibfnamefont {A.~John}\ \bibnamefont
  {Berlinsky}}, \bibinfo {author} {\bibfnamefont {Junhua}\ \bibnamefont
  {Zhang}}, \ and\ \bibinfo {author} {\bibfnamefont {Catherine}\ \bibnamefont
  {Kallin}},\ }\bibfield  {title} {\enquote {\bibinfo {title} {Spin structure
  factor of the frustrated quantum magnet {Cs}$_{2}${CuCl}$_{4}$},}\ }\href
  {\doibase 10.1103/PhysRevB.73.184403} {\bibfield  {journal} {\bibinfo
  {journal} {Phys. Rev. B}\ }\textbf {\bibinfo {volume} {73}},\ \bibinfo
  {pages} {184403} (\bibinfo {year} {2006})}\BibitemShut {NoStop}%
\bibitem [{\citenamefont {Mourigal}\ \emph {et~al.}(2013)\citenamefont
  {Mourigal}, \citenamefont {Fuhrman}, \citenamefont {Chernyshev},\ and\
  \citenamefont {Zhitomirsky}}]{PhysRevB.88.094407}%
  \BibitemOpen
  \bibfield  {author} {\bibinfo {author} {\bibfnamefont {M.}~\bibnamefont
  {Mourigal}}, \bibinfo {author} {\bibfnamefont {W.~T.}\ \bibnamefont
  {Fuhrman}}, \bibinfo {author} {\bibfnamefont {A.~L.}\ \bibnamefont
  {Chernyshev}}, \ and\ \bibinfo {author} {\bibfnamefont {M.~E.}\ \bibnamefont
  {Zhitomirsky}},\ }\bibfield  {title} {\enquote {\bibinfo {title} {Dynamical
  structure factor of the triangular-lattice antiferromagnet},}\ }\href
  {\doibase 10.1103/PhysRevB.88.094407} {\bibfield  {journal} {\bibinfo
  {journal} {Phys. Rev. B}\ }\textbf {\bibinfo {volume} {88}},\ \bibinfo
  {pages} {094407} (\bibinfo {year} {2013})}\BibitemShut {NoStop}%
\bibitem [{\citenamefont {Sugawara}\ and\ \citenamefont
  {Yamada}(1993)}]{Sugawara_1993}%
  \BibitemOpen
  \bibfield  {author} {\bibinfo {author} {\bibfnamefont {K}~\bibnamefont
  {Sugawara}}\ and\ \bibinfo {author} {\bibfnamefont {I}~\bibnamefont
  {Yamada}},\ }\bibfield  {title} {\enquote {\bibinfo {title} {Raman scattering
  study of the triangular-lattice antiferromagnet {VCl}$_2$},}\ }\href
  {\doibase 10.1088/0953-8984/5/9/027} {\bibfield  {journal} {\bibinfo
  {journal} {Journal of Physics: Condensed Matter}\ }\textbf {\bibinfo {volume}
  {5}},\ \bibinfo {pages} {1427--1436} (\bibinfo {year} {1993})}\BibitemShut
  {NoStop}%
\bibitem [{\citenamefont {Suzuki}\ \emph {et~al.}(1993)\citenamefont {Suzuki},
  \citenamefont {Yamada}, \citenamefont {Kadowaki},\ and\ \citenamefont
  {Takei}}]{Suzuki_1993}%
  \BibitemOpen
  \bibfield  {author} {\bibinfo {author} {\bibfnamefont {M}~\bibnamefont
  {Suzuki}}, \bibinfo {author} {\bibfnamefont {I}~\bibnamefont {Yamada}},
  \bibinfo {author} {\bibfnamefont {H}~\bibnamefont {Kadowaki}}, \ and\
  \bibinfo {author} {\bibfnamefont {F}~\bibnamefont {Takei}},\ }\bibfield
  {title} {\enquote {\bibinfo {title} {A {R}aman scattering investigation of
  the magnetic ordering in the two-dimensional triangular lattice
  antiferromagnet {LiCrO}$_{2}$},}\ }\href {\doibase
  10.1088/0953-8984/5/25/012} {\bibfield  {journal} {\bibinfo  {journal}
  {Journal of Physics: Condensed Matter}\ }\textbf {\bibinfo {volume} {5}},\
  \bibinfo {pages} {4225--4232} (\bibinfo {year} {1993})}\BibitemShut {NoStop}%
\bibitem [{\citenamefont {Aktas}\ \emph {et~al.}(2011)\citenamefont {Aktas},
  \citenamefont {Truong}, \citenamefont {Otani}, \citenamefont {Balakrishnan},
  \citenamefont {Clouter}, \citenamefont {Kimura},\ and\ \citenamefont
  {Quirion}}]{Aktas_2011}%
  \BibitemOpen
  \bibfield  {author} {\bibinfo {author} {\bibfnamefont {O}~\bibnamefont
  {Aktas}}, \bibinfo {author} {\bibfnamefont {K~D}\ \bibnamefont {Truong}},
  \bibinfo {author} {\bibfnamefont {T}~\bibnamefont {Otani}}, \bibinfo {author}
  {\bibfnamefont {G}~\bibnamefont {Balakrishnan}}, \bibinfo {author}
  {\bibfnamefont {M~J}\ \bibnamefont {Clouter}}, \bibinfo {author}
  {\bibfnamefont {T}~\bibnamefont {Kimura}}, \ and\ \bibinfo {author}
  {\bibfnamefont {G}~\bibnamefont {Quirion}},\ }\bibfield  {title} {\enquote
  {\bibinfo {title} {Raman scattering study of delafossite magnetoelectric
  multiferroic compounds: {CuFeO}$_{2}$ and {CuCrO}$_{2}$},}\ }\href {\doibase
  10.1088/0953-8984/24/3/036003} {\bibfield  {journal} {\bibinfo  {journal}
  {Journal of Physics: Condensed Matter}\ }\textbf {\bibinfo {volume} {24}},\
  \bibinfo {pages} {036003} (\bibinfo {year} {2011})}\BibitemShut {NoStop}%
\bibitem [{\citenamefont {Wulferding}\ \emph {et~al.}(2012)\citenamefont
  {Wulferding}, \citenamefont {Choi}, \citenamefont {Lemmens}, \citenamefont
  {Ponomaryov}, \citenamefont {van Tol}, \citenamefont {Islam}, \citenamefont
  {Toth},\ and\ \citenamefont {Lake}}]{Wulferding_2012}%
  \BibitemOpen
  \bibfield  {author} {\bibinfo {author} {\bibfnamefont {Dirk}\ \bibnamefont
  {Wulferding}}, \bibinfo {author} {\bibfnamefont {Kwang-Yong}\ \bibnamefont
  {Choi}}, \bibinfo {author} {\bibfnamefont {Peter}\ \bibnamefont {Lemmens}},
  \bibinfo {author} {\bibfnamefont {Alexey~N}\ \bibnamefont {Ponomaryov}},
  \bibinfo {author} {\bibfnamefont {Johan}\ \bibnamefont {van Tol}}, \bibinfo
  {author} {\bibfnamefont {A~T M~Nazmul}\ \bibnamefont {Islam}}, \bibinfo
  {author} {\bibfnamefont {Sandor}\ \bibnamefont {Toth}}, \ and\ \bibinfo
  {author} {\bibfnamefont {Bella}\ \bibnamefont {Lake}},\ }\bibfield  {title}
  {\enquote {\bibinfo {title} {Softened magnetic excitations in the s= 3/2
  distorted triangular antiferromagnet $\alpha$-{CaCr}$_2${O}$_4$},}\ }\href
  {\doibase 10.1088/0953-8984/24/43/435604} {\bibfield  {journal} {\bibinfo
  {journal} {Journal of Physics: Condensed Matter}\ }\textbf {\bibinfo {volume}
  {24}},\ \bibinfo {pages} {435604} (\bibinfo {year} {2012})}\BibitemShut
  {NoStop}%
\bibitem [{\citenamefont {Valentine}\ \emph {et~al.}(2015)\citenamefont
  {Valentine}, \citenamefont {Koohpayeh}, \citenamefont {Mourigal},
  \citenamefont {McQueen}, \citenamefont {Broholm}, \citenamefont {Drichko},
  \citenamefont {Dutton}, \citenamefont {Cava}, \citenamefont {Birol},
  \citenamefont {Das},\ and\ \citenamefont {Fennie}}]{PhysRevB.91.144411}%
  \BibitemOpen
  \bibfield  {author} {\bibinfo {author} {\bibfnamefont {Michael~E.}\
  \bibnamefont {Valentine}}, \bibinfo {author} {\bibfnamefont {Seyed}\
  \bibnamefont {Koohpayeh}}, \bibinfo {author} {\bibfnamefont {Martin}\
  \bibnamefont {Mourigal}}, \bibinfo {author} {\bibfnamefont {Tyrel~M.}\
  \bibnamefont {McQueen}}, \bibinfo {author} {\bibfnamefont {Collin}\
  \bibnamefont {Broholm}}, \bibinfo {author} {\bibfnamefont {Natalia}\
  \bibnamefont {Drichko}}, \bibinfo {author} {\bibfnamefont {Si\^an~E.}\
  \bibnamefont {Dutton}}, \bibinfo {author} {\bibfnamefont {Robert~J.}\
  \bibnamefont {Cava}}, \bibinfo {author} {\bibfnamefont {Turan}\ \bibnamefont
  {Birol}}, \bibinfo {author} {\bibfnamefont {Hena}\ \bibnamefont {Das}}, \
  and\ \bibinfo {author} {\bibfnamefont {Craig~J.}\ \bibnamefont {Fennie}},\
  }\bibfield  {title} {\enquote {\bibinfo {title} {Raman study of magnetic
  excitations and magnetoelastic coupling in
  ${\alpha}$-{SrCr}$_{2}${O}$_{4}$},}\ }\href {\doibase
  10.1103/PhysRevB.91.144411} {\bibfield  {journal} {\bibinfo  {journal} {Phys.
  Rev. B}\ }\textbf {\bibinfo {volume} {91}},\ \bibinfo {pages} {144411}
  (\bibinfo {year} {2015})}\BibitemShut {NoStop}%
\bibitem [{\citenamefont {Drichko}\ \emph {et~al.}(2015)\citenamefont
  {Drichko}, \citenamefont {Hackl},\ and\ \citenamefont
  {Schlueter}}]{PhysRevB.92.161112}%
  \BibitemOpen
  \bibfield  {author} {\bibinfo {author} {\bibfnamefont {Natalia}\ \bibnamefont
  {Drichko}}, \bibinfo {author} {\bibfnamefont {Rudi}\ \bibnamefont {Hackl}}, \
  and\ \bibinfo {author} {\bibfnamefont {John~A.}\ \bibnamefont {Schlueter}},\
  }\bibfield  {title} {\enquote {\bibinfo {title} {Antiferromagnetic
  fluctuations in a quasi-two-dimensional organic superconductor detected by
  raman spectroscopy},}\ }\href {\doibase 10.1103/PhysRevB.92.161112}
  {\bibfield  {journal} {\bibinfo  {journal} {Phys. Rev. B}\ }\textbf {\bibinfo
  {volume} {92}},\ \bibinfo {pages} {161112} (\bibinfo {year}
  {2015})}\BibitemShut {NoStop}%
\bibitem [{\citenamefont {Valentine}\ \emph {et~al.}(2020)\citenamefont
  {Valentine}, \citenamefont {Higo}, \citenamefont {Nambu}, \citenamefont
  {Chaudhuri}, \citenamefont {Wen}, \citenamefont {Broholm}, \citenamefont
  {Nakatsuji},\ and\ \citenamefont {Drichko}}]{2005.06073}%
  \BibitemOpen
  \bibfield  {author} {\bibinfo {author} {\bibfnamefont {Michael~E.}\
  \bibnamefont {Valentine}}, \bibinfo {author} {\bibfnamefont {Tomoya}\
  \bibnamefont {Higo}}, \bibinfo {author} {\bibfnamefont {Yusuke}\ \bibnamefont
  {Nambu}}, \bibinfo {author} {\bibfnamefont {Dipanjan}\ \bibnamefont
  {Chaudhuri}}, \bibinfo {author} {\bibfnamefont {Jiajia}\ \bibnamefont {Wen}},
  \bibinfo {author} {\bibfnamefont {Collin}\ \bibnamefont {Broholm}}, \bibinfo
  {author} {\bibfnamefont {Satoru}\ \bibnamefont {Nakatsuji}}, \ and\ \bibinfo
  {author} {\bibfnamefont {Natalia}\ \bibnamefont {Drichko}},\ }\href@noop {}
  {\enquote {\bibinfo {title} {Impact of the lattice on magnetic properties and
  possible spin nematicity in the s=1 triangular antiferromagnet
  {NiGa}$_2${S}$_4$},}\ } (\bibinfo {year} {2020}),\ \Eprint
  {http://arxiv.org/abs/2005.06073} {arXiv:2005.06073 [cond-mat.str-el]}
  \BibitemShut {NoStop}%
\bibitem [{\citenamefont {Vernay}\ \emph {et~al.}(2007)\citenamefont {Vernay},
  \citenamefont {Devereaux},\ and\ \citenamefont {Gingras}}]{Vernay_2007}%
  \BibitemOpen
  \bibfield  {author} {\bibinfo {author} {\bibfnamefont {F}~\bibnamefont
  {Vernay}}, \bibinfo {author} {\bibfnamefont {T~P}\ \bibnamefont {Devereaux}},
  \ and\ \bibinfo {author} {\bibfnamefont {M~J~P}\ \bibnamefont {Gingras}},\
  }\bibfield  {title} {\enquote {\bibinfo {title} {Raman scattering for
  triangular lattices spin-1/2 heisenberg antiferromagnets},}\ }\href {\doibase
  10.1088/0953-8984/19/14/145243} {\bibfield  {journal} {\bibinfo  {journal}
  {Journal of Physics: Condensed Matter}\ }\textbf {\bibinfo {volume} {19}},\
  \bibinfo {pages} {145243} (\bibinfo {year} {2007})}\BibitemShut {NoStop}%
\bibitem [{\citenamefont {Perkins}\ and\ \citenamefont
  {Brenig}(2008)}]{PhysRevB.77.174412}%
  \BibitemOpen
  \bibfield  {author} {\bibinfo {author} {\bibfnamefont {Natalia}\ \bibnamefont
  {Perkins}}\ and\ \bibinfo {author} {\bibfnamefont {Wolfram}\ \bibnamefont
  {Brenig}},\ }\bibfield  {title} {\enquote {\bibinfo {title} {Raman scattering
  in a heisenberg $s=\frac{1}{2}$ antiferromagnet on the triangular lattice},}\
  }\href {\doibase 10.1103/PhysRevB.77.174412} {\bibfield  {journal} {\bibinfo
  {journal} {Phys. Rev. B}\ }\textbf {\bibinfo {volume} {77}},\ \bibinfo
  {pages} {174412} (\bibinfo {year} {2008})}\BibitemShut {NoStop}%
\bibitem [{\citenamefont {Perkins}\ \emph {et~al.}(2013)\citenamefont
  {Perkins}, \citenamefont {Chern},\ and\ \citenamefont
  {Brenig}}]{PhysRevB.87.174423}%
  \BibitemOpen
  \bibfield  {author} {\bibinfo {author} {\bibfnamefont {Natalia~B.}\
  \bibnamefont {Perkins}}, \bibinfo {author} {\bibfnamefont {Gia-Wei}\
  \bibnamefont {Chern}}, \ and\ \bibinfo {author} {\bibfnamefont {Wolfram}\
  \bibnamefont {Brenig}},\ }\bibfield  {title} {\enquote {\bibinfo {title}
  {Raman scattering in a heisenberg $s=\frac{1}{2}$ antiferromagnet on the
  anisotropic triangular lattice},}\ }\href {\doibase
  10.1103/PhysRevB.87.174423} {\bibfield  {journal} {\bibinfo  {journal} {Phys.
  Rev. B}\ }\textbf {\bibinfo {volume} {87}},\ \bibinfo {pages} {174423}
  (\bibinfo {year} {2013})}\BibitemShut {NoStop}%
\bibitem [{\citenamefont {Jin}\ \emph {et~al.}(2019)\citenamefont {Jin},
  \citenamefont {Luo}, \citenamefont {Datta},\ and\ \citenamefont
  {Yao}}]{PhysRevB.100.054410}%
  \BibitemOpen
  \bibfield  {author} {\bibinfo {author} {\bibfnamefont {Shangjian}\
  \bibnamefont {Jin}}, \bibinfo {author} {\bibfnamefont {Cheng}\ \bibnamefont
  {Luo}}, \bibinfo {author} {\bibfnamefont {Trinanjan}\ \bibnamefont {Datta}},
  \ and\ \bibinfo {author} {\bibfnamefont {Dao-Xin}\ \bibnamefont {Yao}},\
  }\bibfield  {title} {\enquote {\bibinfo {title} {Torque equilibrium spin wave
  theory study of anisotropy and dzyaloshinskii-moriya interaction effects on
  the indirect $k$-edge rixs spectrum of a triangular lattice
  antiferromagnet},}\ }\href {\doibase 10.1103/PhysRevB.100.054410} {\bibfield
  {journal} {\bibinfo  {journal} {Phys. Rev. B}\ }\textbf {\bibinfo {volume}
  {100}},\ \bibinfo {pages} {054410} (\bibinfo {year} {2019})}\BibitemShut
  {NoStop}%
\bibitem [{\citenamefont {Fleury}\ and\ \citenamefont
  {Loudon}(1968)}]{PhysRev.166.514}%
  \BibitemOpen
  \bibfield  {author} {\bibinfo {author} {\bibfnamefont {P.~A.}\ \bibnamefont
  {Fleury}}\ and\ \bibinfo {author} {\bibfnamefont {R.}~\bibnamefont
  {Loudon}},\ }\bibfield  {title} {\enquote {\bibinfo {title} {Scattering of
  light by one- and two-magnon excitations},}\ }\href {\doibase
  10.1103/PhysRev.166.514} {\bibfield  {journal} {\bibinfo  {journal} {Phys.
  Rev.}\ }\textbf {\bibinfo {volume} {166}},\ \bibinfo {pages} {514--530}
  (\bibinfo {year} {1968})}\BibitemShut {NoStop}%
\bibitem [{\citenamefont {Weihong}\ \emph {et~al.}(1999)\citenamefont
  {Weihong}, \citenamefont {McKenzie},\ and\ \citenamefont
  {Singh}}]{PhysRevB.59.14367}%
  \BibitemOpen
  \bibfield  {author} {\bibinfo {author} {\bibfnamefont {Zheng}\ \bibnamefont
  {Weihong}}, \bibinfo {author} {\bibfnamefont {Ross~H.}\ \bibnamefont
  {McKenzie}}, \ and\ \bibinfo {author} {\bibfnamefont {Rajiv R.~P.}\
  \bibnamefont {Singh}},\ }\bibfield  {title} {\enquote {\bibinfo {title}
  {Phase diagram for a class of spin-$\frac{1}{2}$ heisenberg models
  interpolating between the square-lattice, the triangular-lattice, and the
  linear-chain limits},}\ }\href {\doibase 10.1103/PhysRevB.59.14367}
  {\bibfield  {journal} {\bibinfo  {journal} {Phys. Rev. B}\ }\textbf {\bibinfo
  {volume} {59}},\ \bibinfo {pages} {14367--14375} (\bibinfo {year}
  {1999})}\BibitemShut {NoStop}%
\bibitem [{\citenamefont {Susuki}\ \emph {et~al.}(2013)\citenamefont {Susuki},
  \citenamefont {Kurita}, \citenamefont {Tanaka}, \citenamefont {Nojiri},
  \citenamefont {Matsuo}, \citenamefont {Kindo},\ and\ \citenamefont
  {Tanaka}}]{PhysRevLett.110.267201}%
  \BibitemOpen
  \bibfield  {author} {\bibinfo {author} {\bibfnamefont {Takuya}\ \bibnamefont
  {Susuki}}, \bibinfo {author} {\bibfnamefont {Nobuyuki}\ \bibnamefont
  {Kurita}}, \bibinfo {author} {\bibfnamefont {Takuya}\ \bibnamefont {Tanaka}},
  \bibinfo {author} {\bibfnamefont {Hiroyuki}\ \bibnamefont {Nojiri}}, \bibinfo
  {author} {\bibfnamefont {Akira}\ \bibnamefont {Matsuo}}, \bibinfo {author}
  {\bibfnamefont {Koichi}\ \bibnamefont {Kindo}}, \ and\ \bibinfo {author}
  {\bibfnamefont {Hidekazu}\ \bibnamefont {Tanaka}},\ }\bibfield  {title}
  {\enquote {\bibinfo {title} {Magnetization process and collective excitations
  in the $s\mathbf{=}1/2$ triangular-lattice heisenberg antiferromagnet
  {Ba}$_3${CoSb}$_2${O}$_9$},}\ }\href {\doibase
  10.1103/PhysRevLett.110.267201} {\bibfield  {journal} {\bibinfo  {journal}
  {Phys. Rev. Lett.}\ }\textbf {\bibinfo {volume} {110}},\ \bibinfo {pages}
  {267201} (\bibinfo {year} {2013})}\BibitemShut {NoStop}%
\bibitem [{\citenamefont {Frontzek}\ \emph {et~al.}(2011)\citenamefont
  {Frontzek}, \citenamefont {Haraldsen}, \citenamefont {Podlesnyak},
  \citenamefont {Matsuda}, \citenamefont {Christianson}, \citenamefont
  {Fishman}, \citenamefont {Sefat}, \citenamefont {Qiu}, \citenamefont
  {Copley}, \citenamefont {Barilo}, \citenamefont {Shiryaev},\ and\
  \citenamefont {Ehlers}}]{PhysRevB.84.094448}%
  \BibitemOpen
  \bibfield  {author} {\bibinfo {author} {\bibfnamefont {M.}~\bibnamefont
  {Frontzek}}, \bibinfo {author} {\bibfnamefont {J.~T.}\ \bibnamefont
  {Haraldsen}}, \bibinfo {author} {\bibfnamefont {A.}~\bibnamefont
  {Podlesnyak}}, \bibinfo {author} {\bibfnamefont {M.}~\bibnamefont {Matsuda}},
  \bibinfo {author} {\bibfnamefont {A.~D.}\ \bibnamefont {Christianson}},
  \bibinfo {author} {\bibfnamefont {R.~S.}\ \bibnamefont {Fishman}}, \bibinfo
  {author} {\bibfnamefont {A.~S.}\ \bibnamefont {Sefat}}, \bibinfo {author}
  {\bibfnamefont {Y.}~\bibnamefont {Qiu}}, \bibinfo {author} {\bibfnamefont
  {J.~R.~D.}\ \bibnamefont {Copley}}, \bibinfo {author} {\bibfnamefont
  {S.}~\bibnamefont {Barilo}}, \bibinfo {author} {\bibfnamefont {S.~V.}\
  \bibnamefont {Shiryaev}}, \ and\ \bibinfo {author} {\bibfnamefont
  {G.}~\bibnamefont {Ehlers}},\ }\bibfield  {title} {\enquote {\bibinfo {title}
  {Magnetic excitations in the geometric frustrated multiferroic
  {CuCrO}$_2$},}\ }\href {\doibase 10.1103/PhysRevB.84.094448} {\bibfield
  {journal} {\bibinfo  {journal} {Phys. Rev. B}\ }\textbf {\bibinfo {volume}
  {84}},\ \bibinfo {pages} {094448} (\bibinfo {year} {2011})}\BibitemShut
  {NoStop}%
\bibitem [{\citenamefont {Zvyagin}\ \emph {et~al.}(2014)\citenamefont
  {Zvyagin}, \citenamefont {Kamenskyi}, \citenamefont {Ozerov}, \citenamefont
  {Wosnitza}, \citenamefont {Ikeda}, \citenamefont {Fujita}, \citenamefont
  {Hagiwara}, \citenamefont {Smirnov}, \citenamefont {Soldatov}, \citenamefont
  {Shapiro}, \citenamefont {Krzystek}, \citenamefont {Hu}, \citenamefont {Ryu},
  \citenamefont {Petrovic},\ and\ \citenamefont
  {Zhitomirsky}}]{PhysRevLett.112.077206}%
  \BibitemOpen
  \bibfield  {author} {\bibinfo {author} {\bibfnamefont {S.~A.}\ \bibnamefont
  {Zvyagin}}, \bibinfo {author} {\bibfnamefont {D.}~\bibnamefont {Kamenskyi}},
  \bibinfo {author} {\bibfnamefont {M.}~\bibnamefont {Ozerov}}, \bibinfo
  {author} {\bibfnamefont {J.}~\bibnamefont {Wosnitza}}, \bibinfo {author}
  {\bibfnamefont {M.}~\bibnamefont {Ikeda}}, \bibinfo {author} {\bibfnamefont
  {T.}~\bibnamefont {Fujita}}, \bibinfo {author} {\bibfnamefont
  {M.}~\bibnamefont {Hagiwara}}, \bibinfo {author} {\bibfnamefont {A.~I.}\
  \bibnamefont {Smirnov}}, \bibinfo {author} {\bibfnamefont {T.~A.}\
  \bibnamefont {Soldatov}}, \bibinfo {author} {\bibfnamefont {A.~Ya.}\
  \bibnamefont {Shapiro}}, \bibinfo {author} {\bibfnamefont {J.}~\bibnamefont
  {Krzystek}}, \bibinfo {author} {\bibfnamefont {R.}~\bibnamefont {Hu}},
  \bibinfo {author} {\bibfnamefont {H.}~\bibnamefont {Ryu}}, \bibinfo {author}
  {\bibfnamefont {C.}~\bibnamefont {Petrovic}}, \ and\ \bibinfo {author}
  {\bibfnamefont {M.~E.}\ \bibnamefont {Zhitomirsky}},\ }\bibfield  {title}
  {\enquote {\bibinfo {title} {Direct determination of exchange parameters in
  {Cs}$_2${CuBr}$_4$ and {Cs}$_2${CuCl}$_4$: High-field electron-spin-resonance
  studies},}\ }\href {\doibase 10.1103/PhysRevLett.112.077206} {\bibfield
  {journal} {\bibinfo  {journal} {Phys. Rev. Lett.}\ }\textbf {\bibinfo
  {volume} {112}},\ \bibinfo {pages} {077206} (\bibinfo {year}
  {2014})}\BibitemShut {NoStop}%
\bibitem [{\citenamefont {Zvyagin}\ \emph {et~al.}(2015)\citenamefont
  {Zvyagin}, \citenamefont {Ozerov}, \citenamefont {Kamenskyi}, \citenamefont
  {Wosnitza}, \citenamefont {Krzystek}, \citenamefont {Yoshizawa},
  \citenamefont {Hagiwara}, \citenamefont {Hu}, \citenamefont {Ryu},
  \citenamefont {Petrovic},\ and\ \citenamefont {Zhitomirsky}}]{Zvyagin_2015}%
  \BibitemOpen
  \bibfield  {author} {\bibinfo {author} {\bibfnamefont {S~A}\ \bibnamefont
  {Zvyagin}}, \bibinfo {author} {\bibfnamefont {M}~\bibnamefont {Ozerov}},
  \bibinfo {author} {\bibfnamefont {D}~\bibnamefont {Kamenskyi}}, \bibinfo
  {author} {\bibfnamefont {J}~\bibnamefont {Wosnitza}}, \bibinfo {author}
  {\bibfnamefont {J}~\bibnamefont {Krzystek}}, \bibinfo {author} {\bibfnamefont
  {D}~\bibnamefont {Yoshizawa}}, \bibinfo {author} {\bibfnamefont
  {M}~\bibnamefont {Hagiwara}}, \bibinfo {author} {\bibfnamefont {Rongwei}\
  \bibnamefont {Hu}}, \bibinfo {author} {\bibfnamefont {Hyejin}\ \bibnamefont
  {Ryu}}, \bibinfo {author} {\bibfnamefont {C}~\bibnamefont {Petrovic}}, \ and\
  \bibinfo {author} {\bibfnamefont {M~E}\ \bibnamefont {Zhitomirsky}},\
  }\bibfield  {title} {\enquote {\bibinfo {title} {Magnetic excitations in the
  spin-1/2 triangular-lattice antiferromagnet {Cs}$_2${CuBr}$_4$},}\ }\href
  {\doibase 10.1088/1367-2630/17/11/113059} {\bibfield  {journal} {\bibinfo
  {journal} {New Journal of Physics}\ }\textbf {\bibinfo {volume} {17}},\
  \bibinfo {pages} {113059} (\bibinfo {year} {2015})}\BibitemShut {NoStop}%
\bibitem [{\citenamefont {Park}\ \emph {et~al.}(2003)\citenamefont {Park},
  \citenamefont {Park}, \citenamefont {Jeon}, \citenamefont {Choi},
  \citenamefont {Lee}, \citenamefont {Jo}, \citenamefont {Bewley},
  \citenamefont {McEwen},\ and\ \citenamefont {Perring}}]{PhysRevB.68.104426}%
  \BibitemOpen
  \bibfield  {author} {\bibinfo {author} {\bibfnamefont {Junghwan}\
  \bibnamefont {Park}}, \bibinfo {author} {\bibfnamefont {J.-G.}\ \bibnamefont
  {Park}}, \bibinfo {author} {\bibfnamefont {Gun~Sang}\ \bibnamefont {Jeon}},
  \bibinfo {author} {\bibfnamefont {Han-Yong}\ \bibnamefont {Choi}}, \bibinfo
  {author} {\bibfnamefont {Changhee}\ \bibnamefont {Lee}}, \bibinfo {author}
  {\bibfnamefont {W.}~\bibnamefont {Jo}}, \bibinfo {author} {\bibfnamefont
  {R.}~\bibnamefont {Bewley}}, \bibinfo {author} {\bibfnamefont {K.~A.}\
  \bibnamefont {McEwen}}, \ and\ \bibinfo {author} {\bibfnamefont {T.~G.}\
  \bibnamefont {Perring}},\ }\bibfield  {title} {\enquote {\bibinfo {title}
  {Magnetic ordering and spin-liquid state of {YMnO}$_{3}$},}\ }\href {\doibase
  10.1103/PhysRevB.68.104426} {\bibfield  {journal} {\bibinfo  {journal} {Phys.
  Rev. B}\ }\textbf {\bibinfo {volume} {68}},\ \bibinfo {pages} {104426}
  (\bibinfo {year} {2003})}\BibitemShut {NoStop}%
\bibitem [{\citenamefont {Shirata}\ \emph {et~al.}(2012)\citenamefont
  {Shirata}, \citenamefont {Tanaka}, \citenamefont {Matsuo},\ and\
  \citenamefont {Kindo}}]{PhysRevLett.108.057205}%
  \BibitemOpen
  \bibfield  {author} {\bibinfo {author} {\bibfnamefont {Yutaka}\ \bibnamefont
  {Shirata}}, \bibinfo {author} {\bibfnamefont {Hidekazu}\ \bibnamefont
  {Tanaka}}, \bibinfo {author} {\bibfnamefont {Akira}\ \bibnamefont {Matsuo}},
  \ and\ \bibinfo {author} {\bibfnamefont {Koichi}\ \bibnamefont {Kindo}},\
  }\bibfield  {title} {\enquote {\bibinfo {title} {Experimental realization of
  a spin-$1/2$ triangular-lattice heisenberg antiferromagnet},}\ }\href
  {\doibase 10.1103/PhysRevLett.108.057205} {\bibfield  {journal} {\bibinfo
  {journal} {Phys. Rev. Lett.}\ }\textbf {\bibinfo {volume} {108}},\ \bibinfo
  {pages} {057205} (\bibinfo {year} {2012})}\BibitemShut {NoStop}%
\bibitem [{\citenamefont {Koutroulakis}\ \emph {et~al.}(2015)\citenamefont
  {Koutroulakis}, \citenamefont {Zhou}, \citenamefont {Kamiya}, \citenamefont
  {Thompson}, \citenamefont {Zhou}, \citenamefont {Batista},\ and\
  \citenamefont {Brown}}]{PhysRevB.91.024410}%
  \BibitemOpen
  \bibfield  {author} {\bibinfo {author} {\bibfnamefont {G.}~\bibnamefont
  {Koutroulakis}}, \bibinfo {author} {\bibfnamefont {T.}~\bibnamefont {Zhou}},
  \bibinfo {author} {\bibfnamefont {Y.}~\bibnamefont {Kamiya}}, \bibinfo
  {author} {\bibfnamefont {J.~D.}\ \bibnamefont {Thompson}}, \bibinfo {author}
  {\bibfnamefont {H.~D.}\ \bibnamefont {Zhou}}, \bibinfo {author}
  {\bibfnamefont {C.~D.}\ \bibnamefont {Batista}}, \ and\ \bibinfo {author}
  {\bibfnamefont {S.~E.}\ \bibnamefont {Brown}},\ }\bibfield  {title} {\enquote
  {\bibinfo {title} {Quantum phase diagram of the $s=\frac{1}{2}$
  triangular-lattice antiferromagnet {Ba}$_3${CoSb}$_2${O}$_9$},}\ }\href
  {\doibase 10.1103/PhysRevB.91.024410} {\bibfield  {journal} {\bibinfo
  {journal} {Phys. Rev. B}\ }\textbf {\bibinfo {volume} {91}},\ \bibinfo
  {pages} {024410} (\bibinfo {year} {2015})}\BibitemShut {NoStop}%
\bibitem [{\citenamefont {Yamamoto}\ \emph
  {et~al.}(2014{\natexlab{a}})\citenamefont {Yamamoto}, \citenamefont
  {Marmorini},\ and\ \citenamefont {Danshita}}]{PhysRevLett.112.127203}%
  \BibitemOpen
  \bibfield  {author} {\bibinfo {author} {\bibfnamefont {Daisuke}\ \bibnamefont
  {Yamamoto}}, \bibinfo {author} {\bibfnamefont {Giacomo}\ \bibnamefont
  {Marmorini}}, \ and\ \bibinfo {author} {\bibfnamefont {Ippei}\ \bibnamefont
  {Danshita}},\ }\bibfield  {title} {\enquote {\bibinfo {title} {Quantum phase
  diagram of the triangular-lattice $xxz$ model in a magnetic field},}\ }\href
  {\doibase 10.1103/PhysRevLett.112.127203} {\bibfield  {journal} {\bibinfo
  {journal} {Phys. Rev. Lett.}\ }\textbf {\bibinfo {volume} {112}},\ \bibinfo
  {pages} {127203} (\bibinfo {year} {2014}{\natexlab{a}})}\BibitemShut
  {NoStop}%
\bibitem [{\citenamefont {Yamamoto}\ \emph
  {et~al.}(2014{\natexlab{b}})\citenamefont {Yamamoto}, \citenamefont
  {Marmorini},\ and\ \citenamefont {Danshita}}]{PhysRevLett.112.259901}%
  \BibitemOpen
  \bibfield  {author} {\bibinfo {author} {\bibfnamefont {Daisuke}\ \bibnamefont
  {Yamamoto}}, \bibinfo {author} {\bibfnamefont {Giacomo}\ \bibnamefont
  {Marmorini}}, \ and\ \bibinfo {author} {\bibfnamefont {Ippei}\ \bibnamefont
  {Danshita}},\ }\bibfield  {title} {\enquote {\bibinfo {title} {Erratum:
  Quantum phase diagram of the triangular-lattice $xxz$ model in a magnetic
  field [phys. rev. lett. 112, 127203 (2014)]},}\ }\href {\doibase
  10.1103/PhysRevLett.112.259901} {\bibfield  {journal} {\bibinfo  {journal}
  {Phys. Rev. Lett.}\ }\textbf {\bibinfo {volume} {112}},\ \bibinfo {pages}
  {259901} (\bibinfo {year} {2014}{\natexlab{b}})}\BibitemShut {NoStop}%
\bibitem [{\citenamefont {Yamamoto}\ \emph {et~al.}(2015)\citenamefont
  {Yamamoto}, \citenamefont {Marmorini},\ and\ \citenamefont
  {Danshita}}]{PhysRevLett.114.027201}%
  \BibitemOpen
  \bibfield  {author} {\bibinfo {author} {\bibfnamefont {Daisuke}\ \bibnamefont
  {Yamamoto}}, \bibinfo {author} {\bibfnamefont {Giacomo}\ \bibnamefont
  {Marmorini}}, \ and\ \bibinfo {author} {\bibfnamefont {Ippei}\ \bibnamefont
  {Danshita}},\ }\bibfield  {title} {\enquote {\bibinfo {title} {Microscopic
  model calculations for the magnetization process of layered
  triangular-lattice quantum antiferromagnets},}\ }\href {\doibase
  10.1103/PhysRevLett.114.027201} {\bibfield  {journal} {\bibinfo  {journal}
  {Phys. Rev. Lett.}\ }\textbf {\bibinfo {volume} {114}},\ \bibinfo {pages}
  {027201} (\bibinfo {year} {2015})}\BibitemShut {NoStop}%
\bibitem [{\citenamefont {Liu}\ \emph {et~al.}(2017)\citenamefont {Liu},
  \citenamefont {Zhang}, \citenamefont {Zhang}, \citenamefont {Yu},\ and\
  \citenamefont {Wang}}]{PhysRevB.95.104431}%
  \BibitemOpen
  \bibfield  {author} {\bibinfo {author} {\bibfnamefont {Changle}\ \bibnamefont
  {Liu}}, \bibinfo {author} {\bibfnamefont {Anmin}\ \bibnamefont {Zhang}},
  \bibinfo {author} {\bibfnamefont {Qingming}\ \bibnamefont {Zhang}}, \bibinfo
  {author} {\bibfnamefont {Rong}\ \bibnamefont {Yu}}, \ and\ \bibinfo {author}
  {\bibfnamefont {Xiaoqun}\ \bibnamefont {Wang}},\ }\bibfield  {title}
  {\enquote {\bibinfo {title} {Spin-wave approach to the two-magnon raman
  scattering in a
  ${J}_{1x}\text{\ensuremath{-}}{J}_{1y}\text{\ensuremath{-}}{J}_{2}\text{\ensuremath{-}}{J}_{c}$
  antiferromagnetic {H}eisenberg model},}\ }\href {\doibase
  10.1103/PhysRevB.95.104431} {\bibfield  {journal} {\bibinfo  {journal} {Phys.
  Rev. B}\ }\textbf {\bibinfo {volume} {95}},\ \bibinfo {pages} {104431}
  (\bibinfo {year} {2017})}\BibitemShut {NoStop}%
\bibitem [{\citenamefont {Heitler}(1954)}]{Heitler1954THE}%
  \BibitemOpen
  \bibfield  {author} {\bibinfo {author} {\bibfnamefont {W.}~\bibnamefont
  {Heitler}},\ }\href@noop {} {\emph {\bibinfo {title} {The Quantum Theory of
  Radiation}}}\ (\bibinfo  {publisher} {S. L.},\ \bibinfo {year}
  {1954})\BibitemShut {NoStop}%
\bibitem [{\citenamefont {Loudon}(1964)}]{Loudon1964The}%
  \BibitemOpen
  \bibfield  {author} {\bibinfo {author} {\bibfnamefont {Rodney}\ \bibnamefont
  {Loudon}},\ }\bibfield  {title} {\enquote {\bibinfo {title} {The {R}aman
  effect in crystals},}\ }\href {\doibase 10.1080/00018736400101051} {\bibfield
   {journal} {\bibinfo  {journal} {Advances in Physics}\ }\textbf {\bibinfo
  {volume} {13}},\ \bibinfo {pages} {423--482} (\bibinfo {year}
  {1964})}\BibitemShut {NoStop}%
\bibitem [{\citenamefont {Moriya}(1967)}]{1967490}%
  \BibitemOpen
  \bibfield  {author} {\bibinfo {author} {\bibfnamefont {T{o}ru}\ \bibnamefont
  {Moriya}},\ }\bibfield  {title} {\enquote {\bibinfo {title} {Theory of light
  scattering by magnetic crystals},}\ }\href {\doibase 10.1143/JPSJ.23.490}
  {\bibfield  {journal} {\bibinfo  {journal} {Journal of the Physical Society
  of Japan}\ }\textbf {\bibinfo {volume} {23}},\ \bibinfo {pages} {490--500}
  (\bibinfo {year} {1967})}\BibitemShut {NoStop}%
\bibitem [{Note1()}]{Note1}%
  \BibitemOpen
  \bibinfo {note} {Note, in our earlier publication Jin~\protect \emph {et
  al.}~\cite {PhysRevB.100.054410}, there was a typographical error in the
  reported RIXS scattering operator expression. The expression missed the DM
  interaction term which was considered in our study. The correct reported form
  of the RIXS operator expression should be $\protect \mathcal {R}_\protect
  \mathbf {q}=\DOTSB \sum@ \slimits@ _{i,\protect \bm {\delta }}e^{i\protect
  \mathbf {q}\cdot \protect \mathbf {r}_i}[J_{i\protect \bm {\delta }}\protect
  \mathbf {S}_i\cdot \protect \mathbf {S}_{i+\protect \bm {\delta }}-\protect
  \mathbf {D}_{\protect \bm {\delta }}\cdot (\protect \mathbf {S}_i\times
  \protect \mathbf {S}_{i+\protect \bm {\delta }})]$.}\BibitemShut {Stop}%
\bibitem [{\citenamefont {Luo}\ \emph {et~al.}(2015)\citenamefont {Luo},
  \citenamefont {Datta}, \citenamefont {Huang},\ and\ \citenamefont
  {Yao}}]{PhysRevB.92.035109}%
  \BibitemOpen
  \bibfield  {author} {\bibinfo {author} {\bibfnamefont {Cheng}\ \bibnamefont
  {Luo}}, \bibinfo {author} {\bibfnamefont {Trinanjan}\ \bibnamefont {Datta}},
  \bibinfo {author} {\bibfnamefont {Zengye}\ \bibnamefont {Huang}}, \ and\
  \bibinfo {author} {\bibfnamefont {Dao-Xin}\ \bibnamefont {Yao}},\ }\bibfield
  {title} {\enquote {\bibinfo {title} {Signatures of indirect $k$-edge resonant
  inelastic x-ray scattering on magnetic excitations in a triangular-lattice
  antiferromagnet},}\ }\href {\doibase 10.1103/PhysRevB.92.035109} {\bibfield
  {journal} {\bibinfo  {journal} {Phys. Rev. B}\ }\textbf {\bibinfo {volume}
  {92}},\ \bibinfo {pages} {035109} (\bibinfo {year} {2015})}\BibitemShut
  {NoStop}%
\end{thebibliography}%
\end{document}